\documentclass[onecolumn,11pt,amssymb,amsmath,preprintnumbers,nofootinbib,superscriptaddress,aps]{revtex4-2}
\usepackage{slashed} 
\usepackage{graphicx}
\usepackage{feynmp}
\usepackage{amssymb}
\usepackage{array}
\usepackage{amsmath}
\usepackage{soul} 
\usepackage{multirow}
\usepackage{comment}
\usepackage{xcolor}
\usepackage{mathtools}
\usepackage{appendix}
\usepackage[colorlinks=true
,urlcolor=blue
,anchorcolor=blue
,citecolor=blue 
,filecolor=blue
,linkcolor=blue
,menucolor=blue
,linktocpage=true
,pdfproducer=medialab
,pdfa=true
]{hyperref}
\usepackage[nameinlink,capitalize]{cleveref} 

\newcommand{\beq}{\begin{equation}}
\newcommand{\eeq}{\end{equation}}
\newcommand{\bea}{\begin{eqnarray}}
\newcommand{\eea}{\end{eqnarray}}

\newcommand{\brr}[1]{\left(#1\right)}
\newcommand{\srr}[1]{\left[#1\right]}

\begin{document}

 
\begin{flushright}
\hfill UMN-TH-4220/23
\end{flushright}

\title{Top Yukawa Coupling Determination at High Energy Muon Collider}

\author{Zhen Liu}
\affiliation{\it School of Physics and Astronomy, University of Minnesota, Minneapolis, MN, 55455, USA}
\author{Kun-Feng Lyu}
\affiliation{\it School of Physics and Astronomy, University of Minnesota, Minneapolis, MN, 55455, USA}
\author{Ishmam Mahbub}
\affiliation{\it School of Physics and Astronomy, University of Minnesota, Minneapolis, MN, 55455, USA}
\author{Lian-Tao Wang}
\affiliation{\it Department of Physics, University of Chicago, Chicago, IL, 60637, USA}
\affiliation{\it Enrico Fermi Institute, University of Chicago, Chicago, IL 60637, USA}
\affiliation{\it Kavli Institute for Cosmological Physics, University of Chicago, Chicago, IL 60637, USA}
\begin{abstract}
The Top Yukawa coupling profoundly influences several core mysteries linked to the electroweak scale and the Higgs boson. We study the feasibility of measuring the Top Yukawa coupling at high-energy muon colliders by examining the high-energy dynamics of the weak boson fusion to top quark pair processes. A deviation of the Top Yukawa coupling from the Standard Model would lead to a modified $V V \rightarrow t\bar{t}$ process, violating unitarity at high energy. Our analysis reveals that utilizing a muon collider with a center-of-mass energy of 10 TeV and an integrated luminosity of 10 ab$^{-1}$ allows us to investigate the Top Yukawa coupling with a precision surpassing 1.5\%, more than one order of magnitude better than the precision from $t\bar t h$ channel at muon colliders. This precision represents a notable enhancement compared to the anticipated sensitivities of the High-Luminosity LHC (3.4\%) and those at muon colliders derived from the $t\bar{t} H$ process.
\end{abstract}

\maketitle
\tableofcontents

\section{Introduction} 
Following the discovery of the Higgs boson in 2012 \cite{ATLAS:2012yve, CMS:2012qbp}, the elucidation of its properties, particularly its interactions with other Standard Model (SM) fields, has become one of the top priorities of particle physics. Substantial efforts at the LHC have focused on quantifying Higgs interactions with fermions, gauge bosons, and self-couplings. Nonetheless, precision measurements at hadron colliders face constraints due to substantial QCD backgrounds. Muon colliders (MuCs) have emerged as an exciting venue for high-precision Higgs exploration. Recent research accentuates their potential, combining the precision of lepton colliders with high center-of-mass energies, thereby allowing for exploration at scales of 10 TeV or higher in an environment with low background~\cite{Boscolo:2018ytm, Delahaye:2019omf,AlAli:2021let,Black:2022cth,Narain:2022qud,Bose:2022obr,Aime:2022flm,schulte:ipac2022-tuizsp2,Zimmermann:2022xbv,deBlas:2022aow,Forslund:2022xjq,Forslund:2023reu}. Admittedly, one inherent challenge for muon colliders is the short lifetime of muons. The high-energy physics community is working on tackling the challenges and the research to enable a future muon collider, in particular, through an ionization cooling scheme.

A crucial physics objective for future colliders is the precise measurement of Higgs couplings.
The Top Yukawa coupling, one of the least constrained parameters in the SM, holds significance for Higgs research, with its deep connection to the profound puzzle of naturalness and Higgs vacuum stability. 
The recent LHC measurement for Top Yukawa is $y_t = 1.16^{+0.24}_{-0.35}$ \cite{CMS:2020djy}.

This study emphasizes the measurement of the Top Yukawa coupling at muon colliders. The dominance of electroweak gauge boson fusion at muon colliders results from the logarithmic growth of the electroweak parton distribution function (PDF) with energy \cite{Kane:1984bb,Costantini:2020stv, Han:2020uid, AlAli:2021let,Han:2021kes,Garosi:2023bvq,Ma:2022vmy,Fornal:2018znf}. We adopt a factorization approach to compute cross-sections, integrating the enhanced collinear splitting with the vector boson hard scattering. Factorization and resummation ensure a cross-section devoid of collinear divergences in high-energy limits. Our initial assumption only considers a Higgs coupling shift $y_t \rightarrow y_t \brr{1 + \delta_{yt}}$ through a single SMEFT operator $\mathcal{O}_y^t = H^\dagger H \bar{Q} \tilde{H} t_R$. Additionally, we examine a scenario involving a heavy singlet vector-like quark.

The paper is structured as follows: \cref{sec:theoretical_frame} introduces our theoretical framework, \cref{sec:channels} details various partonic channels, \cref{sec:result} presents the results, and \cref{sec:conclusion} contains our conclusions.

\section{Theoretical Framework}\label{sec:theoretical_frame}
The Top Yukawa coupling can be directly measured from processes with final states containing Higgs boson, such as $h t \bar{t}  $. One can also access the Top Yukawa coupling with processes with Higgs boson in the intermediate state but not appearing in the external legs. The scattering amplitude of a process with longitudinal gauge bosons as external states, such as $V V \rightarrow t \bar{t}$, could grow with energy if a coupling deviates from the SM. As a result, such an amplitude eventually violates unitarity at a high-energy scale, implying the breakdown of the low-energy description and the appearance of new physics. Similar study on the gauge boson scattering has been carried on in~\cite{Barger:1995cn,Bagger:1993zf,Han:1997nq,Larios:1997ey,Larios:1997dc} to probe the dynamics of electroweak symmetry breaking and various new physics sectors.

In this section, we will first examine the effects of the anomalous Top Yuakawa coupling in the $W_L^+ W_L^- \rightarrow t \bar{t}$ channel and focus on the energy dependence of the scattering amplitude. 
We then discuss some possible UV models that can generate such Top Yukawa coupling, and further discuss the electroweak PDF which would be convoluted with the $V V \rightarrow t \bar{t}$ cross-section to obtain the total signal rate at muon colliders.

\subsection{Perturbative unitarity in the \texorpdfstring{$W_L^+ W_L^- \rightarrow  t \bar{t} $ }  PProcess}

\unitlength = 1mm

\begin{figure}
    \centering
\includegraphics[width=0.9\textwidth]{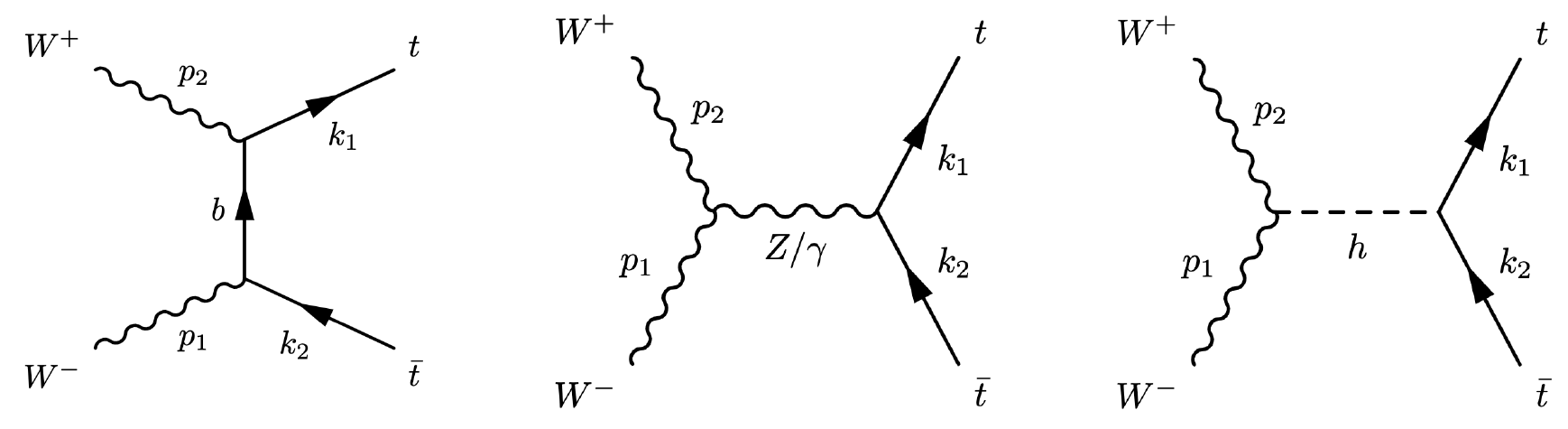}
\caption{The Feynman diagrams for the process $W^+ W^-\rightarrow t \bar{t} $.}
    \label{fig:feyn_diagrams}
\end{figure}
In this section, we consider the channel $W^+_L W^-_L \rightarrow t \Bar{t} $ to illustrate the role of perturbative unitarity in the process. Our analysis of $t \bar{t}$ production in muon colliders involves initial state vector boson WW, ZZ, Z$\gamma$ and $\gamma \gamma$, but we choose $W^+_L W^-_L \rightarrow  t \Bar{t} $ as our example here since it is the dominant channel in our analysis. As shown in \cref{fig:feyn_diagrams}, the Feynman diagrams for the process can be divided into three categories. The first is the $t$-channel diagram via the exchange of the $b$ quark. The second is the $s$-channel diagram mediated by the neutral vector boson, and the last diagram is mediated by the Higgs boson in the $s$-channel. The leading order contribution for the amplitude of the first and second diagrams can be estimated to be $O(E^2/m_W^2)$ by dimensional analysis for $t,\bar{t}$ helicity of $(\pm,\mp)$ and $O(E/m_W)$ for $t,\bar{t}$ helicity of $(\pm,\pm)$. For the case of opposite outgoing helicity of the quarks, once we add the contributions from individual diagrams, the leading $O(E^2/m_W^2)$ behavior of each diagram is canceled through gauge symmetry, leaving a constant term as would be needed for unitarity. While for the case of the same quark helicity, the leading order energy scaling is not eliminated without adding the contribution from the Higgs-mediated diagram. After combining all four diagrams, the energy growth behavior is precisely canceled and perturbative unitarity in this process is restored. Such a cancellation is expected from the Goldstone boson equivalence theorem. In the high energy limit, we have:
\begin{equation}
    \mathcal{M}_{W_L^+ W_L^- \rightarrow t \bar{t}} = \mathcal{M}_{\phi^+ \phi^-\rightarrow  t \bar{t} } \srr{1 + O\brr{\dfrac{m_W^2}{E^2}} }
\end{equation}
The amplitude for $\mathcal{M}_{\phi^+ \phi^-\rightarrow  t \bar{t} }$ in SM is at most a constant. Hence, we do not expect unitarity-violating behavior in the $W^+_L W^-_L \rightarrow  t \Bar{t} $ process.

If the Top Yukawa is shifted,
\begin{equation}
\label{eq:DeltaYt}
    y_t \rightarrow y_t (1 + \delta_{yt}), 
\end{equation}
only the amplitude of the Higgs-mediated diagram will be modified from the SM, and one can show that in the high-energy limit, the total scattering amplitude is
\begin{equation}
\mathcal{M} (W^+ W^- \rightarrow  t \bar{t}) =  \, \frac{m_t }{\nu^2} \delta_{yt} \sqrt{s}  
\quad  \text{for} \quad
\sqrt{s} \gg  m_t .
\label{eq:non-unitary-growth}
\end{equation}

In order to preserve unitarity, or more precisely, to maintain the validity of a weakly coupled EFT, there must be a cutoff. To estimate this cut-off, we can follow the standard partial wave analysis. The J = 0 partial wave amplitude is
\begin{equation}
    a_0 = \frac{1}{32 \pi} \int_{-1}^{1} d \cos \theta |\mathcal{M}|
\end{equation}
Unitarity requires the amplitude $|a_0| \leq 1$. Hence, we obtain the  cutoff as (ignoring the constant term)
\begin{equation}
    \Lambda_{\rm BSM} < \frac{16 \pi v^2}{m_t \, \delta_{yt} }. 
\end{equation}

The current precision on the Top Yukawa coupling $y_t$ is around $\mathcal{O}(10)\%$. Therefore,  the cutoff scale is well above the center of mass energy of a  10 TeV muon collider. We note that the parameterization in \cref{eq:DeltaYt} is (overly) simplified. We can also characterize the deviation in terms of the contribution from new physics. In the following, we describe two approaches to parameterize such contributions. 


\subsection{Dimension-Six Operator}
The most straightforward way to express the deviation of the Top Yukawa coupling from the SM is in terms of the higher dimensional operators in SMEFT~\cite{Grzadkowski:2010es,Jenkins:2013zja,Brivio:2017vri}, generated after integrating out heavy new physics~\cite{Henning:2014wua,Henning:2016lyp}. One can assume the leading EFT operator is $\mathcal{O}_y^t = \brr{H^\dagger H} \bar{Q}_L \tilde{H} t_R $ with the corresponding Wilson coefficient $c_y^t$.\footnote{For simplicity, we consider the $c_y^t$ being real here. One can study consistently defined CP-violation effects in detail in differential decays, which will not affect the inclusive rate that we focus on here.} After Electroweak Symmetry Breaking (EWSB) and combining it with the Top Yukawa coupling term, we get
\begin{equation}
\label{eq:YukSMEFT}
    \mathcal{L} \supset - \srr{ m_t   +  \brr{ \dfrac{m_t}{v} - \dfrac{c_y^t  v^2}{\sqrt{2}} } h - \dfrac{3 c_y^t v}{2 \sqrt{2} } h^2 -   - \dfrac{3 c_y^t }{2 \sqrt{2} } h^3
  } \brr{\bar{t}_L t_R + \bar{t}_R t_L} 
\end{equation}
in the unitary gauge. After imposing the top pole-mass physical boundary condition, the Top Yukawa coupling is shifted by 
\begin{equation}
    \delta_{yt} =  - \dfrac{c^t_y v^3}{\sqrt{2} m_t}.
\end{equation}
The additional terms in \cref{eq:YukSMEFT} are higher dimensional operators describing interactions between the top quark pair and two or three Higgs bosons. Since we are considering the $V V \rightarrow t \bar{t}$ process, only the single Higgs coupling would enter the calculations. Beyond the scenario only the operator $O_y^t$ is turned on, other operators could also enter the process.
Such an assumption, while minimal, may be oversimplifying. In addition to $O_y^t$,  the UV models often generate multiple operators at the same time. Next, we present one simple UV scenario to generate the effective anomalous Yukawa coupling in the IR.

\subsection{Vector-Like Quarks}
Models with Vector-like quarks (VLQs), as a simple extension of SM, are motivated in a broad class of new physics theories (see, e.g., Refs.~\cite{Aguilar-Saavedra:2013qpa,Ellis:2014dza,Alves:2023ufm} and references therein). 
As seen from its name, it is similar to quarks, transforming as a triplet under the SU(3)$_C$ group. Its left and right-handed components carry the same color and electroweak charge. It is possible to write a mass term independent of the Higgs VEV.
Therefore, it has a decoupling limit and it is consistent with the electroweak precision and Higgs precision data.  We can exploit its mixing with the third generation quarks to obtain the effective non-SM Yukawa coupling at low energies.

We start from the simple case that the VLQ is an SM electroweak SU(2) singlet~\cite{Peralta:2017qhu} and has the same quantum numbers as right-handed top quark, $t_R$. Denoting the one flavor VLQ as $T$, and its left and right-handed components are $T_L$ and $T_R$. The relevant Lagrangian is
\begin{equation}
    \mathcal{L} \supset i\bar{T} \slashed{D} T - \brr{ \lambda_1 \bar{Q}_L \tilde{H} t_R + \lambda_2 \bar{Q}_L \tilde{H} T_R  + M_1 \bar{T}_L t_R + M_2 \bar{T}_L T_R } \  .
\end{equation}
However, we can always combine $t_R$ and $T_R$ to get a new right-handed quark, which we will still denote as $T_R$. This field redefinition can remove the mass mixing term between $T_L$ and $t_R$. The resulting Lagrangian is
\begin{equation}
    \mathcal{L} \supset i\bar{T} \slashed{D} T - \brr{ \lambda_t \bar{Q}_L \tilde{H} t_R + \lambda_0 \bar{Q}_L \tilde{H} T_R   + M_0 \bar{T}_L T_R } \  .
\end{equation}
After EWSB we can write down the mass matrix as follows
\begin{equation}
    -\mathcal{L} \supset  
  \begin{pmatrix}  
  \bar{t}_L &   \bar{T}_L
  \end{pmatrix}
  \begin{pmatrix}  
  \lambda_t v/\sqrt{2} & \hspace{0.5cm}  \lambda_0 v/\sqrt{2} \\
  0 & M_0 
  \end{pmatrix}
  \begin{pmatrix}  
  t_R \\  T_R
  \end{pmatrix} + \text{h.c.} 
 = \bar{F}_L M F_R  + \text{h.c.} 
\end{equation}
in which we abbreviate the left-handed and right-handed fermions into the column vectors, $F_L$ and $F_R$, respectively. The mass matrix is denoted as $M$. Note that for fermions, we need separate transformations for the left and right-handed components, $U_L$ and $U_R$, respectively, to diagonalize the mass matrix. We have
\begin{equation}
    U_L M U_R^\dagger = \Lambda = 
  \begin{pmatrix}
      m_t &   \\
      & m_T 
  \end{pmatrix} \quad \text{with} \quad
  U_{L,R} = \begin{pmatrix}
      c_{L,R} & s_{L,R} \\
      -s_{L,R} & c_{L,R}
  \end{pmatrix}
\end{equation}
where $m_t$ and the $M_T$ are the physical masses of the top quark and the VLQ. In the mass basis, the Yukawa matrix can be written as
\begin{equation}
    U_L \begin{pmatrix}
      \lambda_t & \quad \lambda_0   \\
      0 & \quad 0  
  \end{pmatrix} U_R^\dagger = \dfrac{\sqrt{2}}{v} 
  \, U_L \begin{pmatrix}
      1 & \quad 0\\ 0 & \quad 0
  \end{pmatrix} U_L^\dagger \Lambda = 
  \dfrac{\sqrt{2} }{v} \begin{pmatrix}
      c_L^2 m_t & -c_L s_L m_T \\
      -c_L s_L m_T & s_L^2 m_T
  \end{pmatrix} \  .
\end{equation}
On the other hand, using the identity $U_L M M^\dagger U_L^\dagger = \Lambda^2$, we can obtain the relative shift for the Top Yukawa coupling 
\begin{equation}\label{eq:VLQ_dyt_exp}
    \delta_{yt} = -s_L^2 = -\dfrac{\brr{\lambda_t^2 + \lambda_0^2} v^2 /2 - m_t^2}{m_T^2 - m_t^2}.
\end{equation}
In the limit $m_T \gg m_t$, we have $\delta_{yt}\sim v^2/m_T^2$, as expected from the correction induced by a dimension-6 operator. 
Moreover, the fermion field rotation would modify the coupling between the top quark and the gauge field. Making the replacement that $t_L \rightarrow c_L t_L - s_L T_L$, we get
\begin{equation}
\begin{split}
    \dfrac{g}{\sqrt{2} }  W^+_\mu \, \bar{t}_L \gamma^\mu b_L   \rightarrow & \dfrac{g}{\sqrt{2} }  W^+_\mu \, \brr{ c_L \, \bar{t}_L \gamma^\mu b_L  - s_L \, \bar{T}_L \gamma^\mu b_L} \ ,
    \\
    \dfrac{g}{\cos\theta_w} \brr{\dfrac{1}{2} - \dfrac{2}{3} \sin^2\theta_w} Z_\mu \bar{t}_L \gamma^\mu t_L  \rightarrow &
    \dfrac{g}{\cos\theta_w} \brr{\dfrac{1}{2} - \dfrac{2}{3} \sin^2\theta_w}  Z_\mu \\
    &
    \srr{ c_L^2 \bar{t}_L \gamma^\mu t_L -  c_L s_L \brr{\bar{t}_L \gamma^\mu T_L + \bar{T}_L \gamma^\mu t_L} + s_L^2 \bar{T}_L \gamma^\mu T_L }
    \\
     \dfrac{g}{\cos\theta_w} \brr{- \dfrac{2}{3} \sin^2\theta_w} Z_\mu \bar{T}_L \gamma^\mu T_L\rightarrow &  \dfrac{g}{\cos\theta_w} \brr{ - \dfrac{2}{3} \sin^2\theta_w}  Z_\mu
     \\
     &
    \srr{ s_L^2 \bar{t}_L \gamma^\mu t_L +  c_L s_L \brr{\bar{t}_L \gamma^\mu T_L + \bar{T}_L \gamma^\mu t_L} + c_L^2 \bar{T}_L \gamma^\mu T_L }.
\end{split}
\end{equation}
Apart from the modified $Z t \bar{t}$ coupling, there are also new off-diagonal couplings of the form $V-T-t$, with $V$ being the charged and neutral gauge bosons. These couplings would modify the electroweak oblique parameters $T$ and $S$, thus being highly constrained. However, one can always add more vector-like quarks carrying the same charge as the bottom quark to compensate. 
Alternatively, we can integrate out the heavy quark and obtain the corresponding matched EFT Lagrangian. At tree level, we get
\begin{equation}
    \dfrac{\lambda^2_0}{M_0^2} \Bar{Q}_L \Tilde{H} i \slashed{D} \brr{ \Tilde{H}^\dagger Q_L}
    \label{eq:TopKineticOperator}
\end{equation}
After EWSB, this operator leads to the corrections to the top couplings consistent with the mixing effect discussed above at the leading order. The single operator in \cref{eq:TopKineticOperator} can be converted into the operators in Warsaw basis of SMEFT via the equation of motion. We get
 \begin{equation}
     \dfrac{\lambda^2_0}{M_0^2} \srr{ y_t H^\dagger H \, \bar{Q}_L \tilde{H} t_R  
      + \dfrac{1}{4} H^\dagger i \overleftrightarrow{D_\mu} H \, \bar{Q}_L  \gamma^\mu Q_L - \dfrac{1}{4}  H^\dagger \sigma^a i \overleftrightarrow{D_\mu} H \,\bar{Q}_L \sigma^a \gamma^\mu Q_L } 
 \end{equation}
Compared to the coupling modifications discussed in previous sections, apart from the $\mathcal{O}_y^t$ operator, two more operators with correlated coefficients appear and modify the coupling of the charged and neutral current of the third-generation quarks. These correlations have implications when we interpret our results in later sections.

\subsection{Electroweak Gauge Boson PDFs}
Vector Boson Fusion (VBF) production is one of the most important production channels at high-energy colliders. For example, the Higgs boson can be produced by VBF, in addition to the $s$-channel Higgsstrahlung channels at future lepton colliders. As the center of mass energy $\sqrt{s}$ increases, the VBF channel becomes more and more important than the $s$ channel.  For muon colliders with benchmark $\sqrt{s} = 3$ or 10 TeV, the VBF processes dominate the others for Higgs production. In many cases, a high-energy muon collider can be considered a gauge boson collider; the virtual gauge bosons split off the incoming muon and anti-muon beams, then participate in hard scatterings. 

There are a couple of challenges in computing the VBF process accurately at high-energy colliders. 
Splitting a muon into a virtual gauge boson plus remnant lepton is highly forward at high energies, and complexities arise for massive gauge bosons. 
One useful approximation is to treat the nearly on-shell gauge boson as the parton within the incoming muon.
Analogous to the proton case, which can be seen as the collection of separate partons in the QCD perturbative energy scale, the total collision cross section is the sum of the convoluted partonic cross section with the parton distribution function.  A factorization procedure similar to hadron collisions can be applied here. The full VBF process can be decomposed into the muon splitting part and the gauge boson scattering sub-process. 
One can derive the PDF for the virtual gauge bosons through the splitting function. Theoretically, it requires solving the DGLAP equation and matching between the massless splitting and massive gauge bosons. The matching can only be meaningfully done with higher-order calculation information, which is beyond the scope of this study. The gauge boson scattering sub-processes can be convoluted with the PDF to obtain the cross-section. The total cross-section can be expressed as 
\begin{equation}
    \sigma (\mu^+ \mu^- \rightarrow F + X ) \ = \ \int_{\tau_{\rm min}}^{\tau_{\rm max}} d\tau \sum_{ij} \frac{\mathcal{L}_{ij}}{d\tau} \hat{\sigma} (ij \rightarrow F ),
\end{equation}
where $X$ refers to the collinear remnant partons, $F$ stands for the collection of the final state particles from the hard scattering, and $i$, $j$ label the intermediate virtual gauge bosons.
$\tau  =  \hat{s}/{s}$ is the fraction of the partonic center of mass energy. 
The parton luminosity function is defined as
\begin{equation}
    \frac{d \mathcal{L}_{ij}}{d\tau} \ = \ \frac{1}{1+\delta_{ij}} \int_{\tau}^1 \frac{d \xi}{\xi} \srr{ f_i\brr{\xi,\mu_f} f_j\brr{\frac{\tau}{\xi},\mu_f} + i \leftrightarrow j }
\end{equation}
in which $f_i(\xi, \mu_f)$  is the PDF  for the parton $i$ carrying a fraction of the longitudinal momentum of the incoming muon at the factorization scale $\mu_f$. In the following computation, we set both the factorization scale $\mu_f$ and the renormalization scale equal to $\sqrt{\hat{s}}/2 $~\cite{Han:2020uid}. 

The photon PDF is the largest among all gauge bosons due to its huge flux starting from a low energy scale. 
However, photon-initiated channels make little contribution to the signal part of our analysis which comes from the Higgs exchange diagram.
Our main signal channel is from the longitudinal gauge boson initial states, especially $W_L^+ W_L^- \rightarrow t \bar{t}$. The longitudinal gauge boson ($W_L,Z_L$) PDF has no scale dependence at the leading order~\cite{Chen:2016wkt,Bauer:2017isx,Ciafaloni:2001vt}, which can be seen as the remnant from the electroweak symmetry breaking. The suppression from the longitudinal gauge boson PDF can be compensated by the large partonic cross section. These features will be apparent in \cref{sec:channels} and key to our sensitivity analysis of the deviation of Top Yukawa coupling.

\section{Relevant Channels at Muon Collider}\label{sec:channels}

At a high-energy muon collider, the primary production channel of top quark pairs is from the VBF process. In this section, we will compute the SM partonic cross-sections for the processes $V V \rightarrow t \bar{t}$ and then present the results after convolution with the muon PDF. {\tt FeynCalc}\cite{Shtabovenko:2020gxv} was first used to calculate the helicity amplitude and the cross-sections. We then modify the vertices in the SM UFO model files within {\tt MadGraph5\textunderscore aMC$@$NLO}~\cite{Alwall:2014hca,Ruiz:2021tdt}  as cross-check. The gauge boson PDF is computed at the truncated to leading logarithmic order as done in~\cite{AlAli:2021let}.

\subsection{Polarized Partonic Cross Section}\label{subsec:partonic_cross}
\begin{figure}[!ht] 
    \centering
\includegraphics[height=5cm]{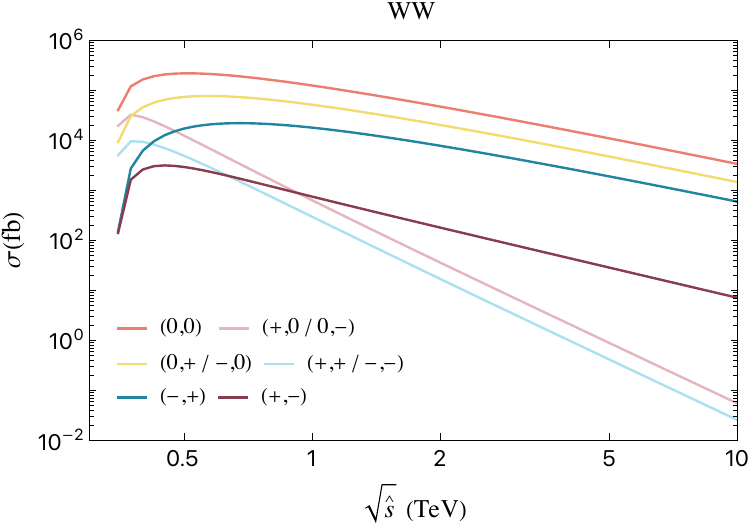}
\hspace{10mm}
\includegraphics[height=5cm]{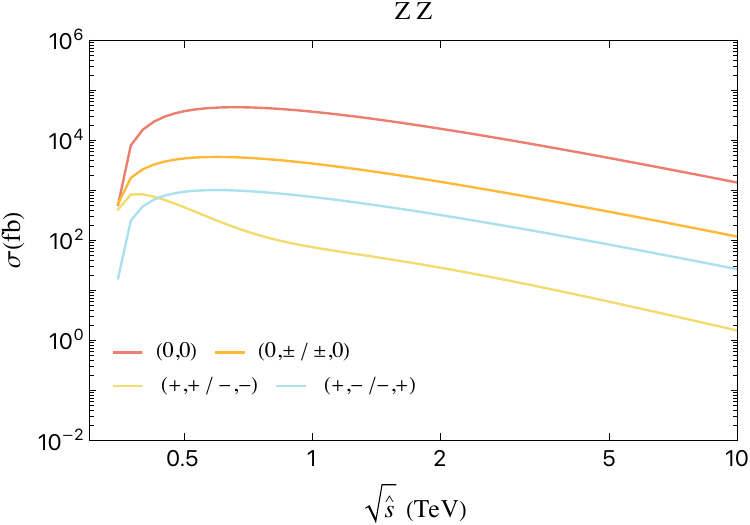}\\ \vspace{2mm}
 \includegraphics[height=5cm]{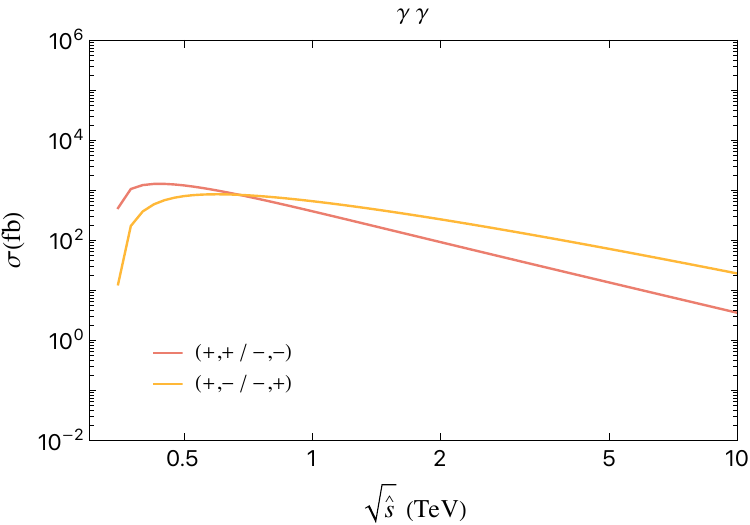}
 \hspace{10mm}
\includegraphics[height=5cm]{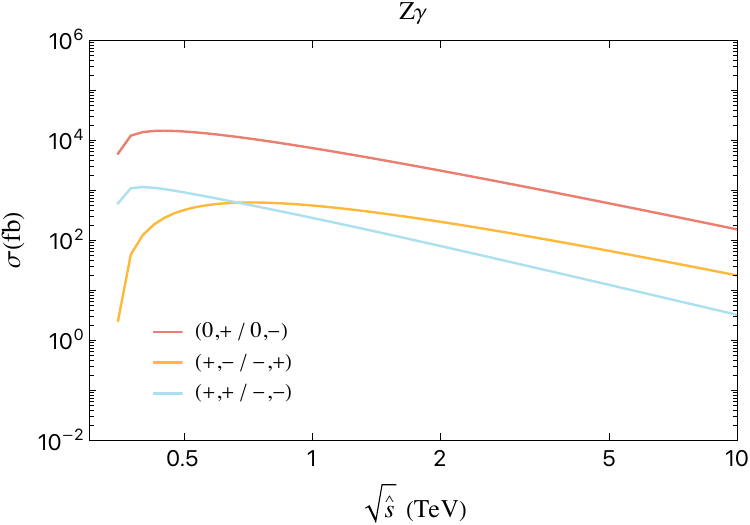}
\caption{The SM partonic cross-section for  $V V \rightarrow t \bar{t}$ as a function of the center of mass energy $\sqrt{\hat{s}}$ with different initial helicities states.}
\label{fig:partonic_distribution}
\end{figure}

We first present the SM predictions of the partonic cross sections in \cref{fig:partonic_distribution}, where all the possible initial polarization states of the vector bosons are separately shown for the range of center-of-mass energies from the $t \bar t$ threshold to 10 TeV. First, let us look at the $WW$ initial states, the most dominant $t\bar{t}$ production channel. The figure shows that the primary channel is the one in which both incoming $W^{\pm}$ are longitudinally polarized. The subdominant contribution to the cross-section occurs for right-handed $W^-$ with longitudinal $W^+$ as well as the CP conjugated combination. In the high energy regime, we observe from the figure that the cross-sections of the channels with initial polarization of $(W^+, W^-)$ to be $(0,0)$, $(0,+)$, $(-,0)$, $(-,+) (+,-)$ scale as $\hat{s}^{-1}$ and the cross-sections of the four other channels scale as $\hat{s}^{-2}$. Therefore, the amplitude squared term does not scale for $(0,0)$, $(0,+)$, $(-,0)$, $(-,+) (+,-)$ and for $(+,0)$, $(0,-)$, $(-,-)$, $(+,+)$ scales as $\hat{s}^{-1}$.
\begin{table}[h]
\begin{tabular}{|c|c|cccc|}
\hline
\multirow{2}{*}{$W^{+}$} & \multirow{2}{*}{$W^{-}$} & \multicolumn{4}{c|}{($t$, $\bar{t}$)}                                                          \\ \cline{3-6} 
                         &                          & \multicolumn{1}{c|}{(+,+)} & \multicolumn{1}{c|}{$(-,+)$} & \multicolumn{1}{c|}{$(+,-)$} & $(-,-)$ \\ \hline
0                        & 0                        & \multicolumn{1}{c|}{$\hat{s}^{-1}$}      & \multicolumn{1}{c|}{$\hat{s}^{0}$}      & \multicolumn{1}{c|}{$\hat{s}^{0}$}      &  $\hat{s}^{-1}$     \\ \hline
+                       & 0                        & \multicolumn{1}{c|}{$\hat{s}^{-2}$}      & \multicolumn{1}{c|}{$\hat{s}^{-1}$}      & \multicolumn{1}{c|}{$\hat{s}^{-1}$}      &  $\hat{s}^{-2}$     \\ \hline
 $-$                        & 0                        & \multicolumn{1}{c|}{$\hat{s}^{-2}$}      & \multicolumn{1}{c|}{$\hat{s}^{-1}$}      & \multicolumn{1}{c|}{$\hat{s}^{-1}$}      &  $\hat{s}^{0}$     \\ \hline
0                        & +                        & \multicolumn{1}{c|}{$\hat{s}^{0}$}      & \multicolumn{1}{c|}{$\hat{s}^{-1}$}      & \multicolumn{1}{c|}{$\hat{s}^{-1}$}      &  $\hat{s}^{-2}$     \\ \hline
+                        & +                        & \multicolumn{1}{c|}{$\hat{s}^{-1}$}      & \multicolumn{1}{c|}{$\hat{s}^{-2}$}      & \multicolumn{1}{c|}{$\hat{s}^{-2}$}      & $\hat{s}^{-3}$      \\ \hline
$-$                        & +                      & \multicolumn{1}{c|}{$\hat{s}^{-1}$}      & \multicolumn{1}{c|}{$\hat{s}^{0}$}      & \multicolumn{1}{c|}{$\hat{s}^{-2}$}      & $\hat{s}^{-1}$      \\ \hline
0                        & $-$                        & \multicolumn{1}{c|}{$\hat{s}^{-2}$}      & \multicolumn{1}{c|}{$\hat{s}^{-1}$}      & \multicolumn{1}{c|}{$\hat{s}^{-1}$}      &  $\hat{s}^{-2}$      \\ \hline
+                        & $-$                        & \multicolumn{1}{c|}{$\hat{s}^{-1}$}      & \multicolumn{1}{c|}{$\hat{s}^{0}$}      & \multicolumn{1}{c|}{$\hat{s}^{-2}$}      &  $\hat{s}^{-1}$     \\ \hline
$-$                        & $-$                        & \multicolumn{1}{c|}{$\hat{s}^{-3}$}      & \multicolumn{1}{c|}{$\hat{s}^{-2}$}      & \multicolumn{1}{c|}{$\hat{s}^{-2}$}      & $\hat{s}^{-1}$      \\ \hline
\end{tabular}
\caption{The high energy $\hat{s}$ scaling of the squared amplitude for $W^+ W^- \rightarrow t \bar{t}$ where helicities of the outgoing top quarks and incoming $W$ bosons are explicitly shown. Here we have ignored the potential additional logarithmic dependence. For the fermion helicities, we use $\pm$ to denote their helicity.
}
\label{tab:WWtt_cross_section}
\end{table}

 To illustrate the scaling mentioned, we show the explicit $\hat{s}$ dependence of the amplitude squared term of all possible helicity configurations for the $W^{+}(h_i) + W^{-}(h_j) \rightarrow  t(h_k) + \bar{t}(h_l)$ processes in \cref{tab:WWtt_cross_section},
where the helicity of the particle is shown in the brackets. Here, $h_i,h_j$ can take value of $1$, 0 or $-1$ while $h_k$ and $h_l$ can be $\pm 1/2$ (labelled as $\pm$ in the table for brevity). 
The tree-level helicity amplitudes are invariant under the CP transformation
\begin{equation}
    W^{+}(h_i) \rightarrow W^{-}(-h_i), \quad
    W^{-}(h_j) \rightarrow W^{+}(-h_j), \quad
    t(h_k) \rightarrow \bar{t}(-h_k), \quad
    \bar{t}(h_l) \rightarrow t(-h_l)
\end{equation}
which is manifest in the table. For example, if one compares the third and fourth row, $(W^+(-),W^-(0))$ is transformed to $(W^+(0),W^-(+))$ and the $\hat{s}$ power is indeed swapped under the CP transformation of the top quark pairs.  

We now focus on the scaling of helicity amplitude for $W_L^+ W_L^- \rightarrow t \bar{t}$ due to its significance as the standard model background as well as the source of the signal. \footnote{$t\bar t$ can also be produced through the $s$-channel process, but it is by far subdominant at high energy muon colliders, and easily distinguishable from the signal process considered here due to its energetic top quark jet, high invariant mass, and low rapidity of the $t\bar t$ system.} Let us first consider $t, \bar{t}$ helicity $(\pm,\pm)$. For the Top Yukawa coupling we are probing, this channel is the largest source of signal in our analysis due to the Higgs exchange diagram. 
Each individual diagram for this helicity scales as $\sqrt{\hat{s}}$ as seen from \cref{eq:a14,eq:a15,eq:a16,eq:a17} and \cref{eq:a26,eq:a27,eq:a28,eq:a29}  in the high energy limit. Once we add all four diagram the leading terms cancels and in SM the amplitude scales as 1/$\sqrt{\hat{s}}$ which is the next leading order term in all these amplitudes. Note that the signal comes only from the Higgs diagram scaling as $\sqrt{\hat{s}}$ and so its interference with SM therefore does not scale.
We elaborate interference and relevant scaling in \cref{sec:ano_top_yukawa}.

The leading contribution of $W^+_L W^-_L \rightarrow  t \bar{t}$ comes from the final quark helicity $ t(+) \bar{t}(-) $ and so we now focus on $t,\bar{t}$ helicity $(\pm, \mp)$ to understand their behavior. As shown by the helicity amplitude in \cref{sec:helicity amplitudes}, the amplitudes of the processes with final quarks helicities $(-,+)$ or $(+,-)$ approach a constant in the high energy limit given by  
\begin{align}
    \mathcal{M}\brr{W^+_L W^-_L \rightarrow t(+) \bar{t}(-) } &=  - \frac{2 m_t^2}{v^2} \cot\dfrac{\theta}{2} + \dfrac{g^{\prime 2}}{3} \sin \theta \ + \mathcal{O} (1/\hat{s}), \label{eq:M-LL+-}\\
    \mathcal{M}\brr{W^+_L W^-_L \rightarrow t(-) \bar{t}(+) } &= \dfrac{3 g^2 + g^{\prime 2}}{12} \sin\theta  + \mathcal{O} (1/\hat{s}) \ . \label{eq:M-LL-+}
\end{align}
where $\theta$ is the angle between $W^+$ and $t$ in the partonic center of mass frame. The leading order behavior of these amplitudes can be understood from the perspective of the Goldstone boson equivalence theorem and from the process $\phi^+ \phi^-\rightarrow t \bar{t}$. Let us first consider $\phi^+ \phi^-\rightarrow t (+) \bar{t}(-)$. At leading order, the $s$-channel process has contribution only from the U(1) field $B_{\mu}$ exchange and no contribution from the $SU(2)_L$ field, which will be of order $O(m_t/\sqrt{\hat{s}})$ for this helicity configuration. The $s$-channel $B_{\mu}$ diagram is proportional to $g^{\prime 2}$ and the Wigner function $d^{1}_{0,1} = \sin\theta/\sqrt{2}$. The process also has a $t$-channel contribution from $b$-quark exchange which will come from terms in Lagrangian of the form $y_t \bar{b}_L t_R \phi^-$ and its hermitian conjugate. This explains why in \cref{eq:M-LL+-}, we observe $t$-channel behavior proportional to $y_t^2$. Now, let us consider the channel $\phi^+ \phi^-\rightarrow t (-) \bar{t}(+)$. In the leading order, the $s$-channel diagram is mediated by $SU(2)_L$ charge neutral vector boson $W_{\mu}^3$ with coupling $g$ and U(1) field $B_{\mu}$ with coupling $g^{\prime}$. Since we assume $y_b = 0$ in our calculation, the leading order $t$-channel contribution is of order $\mathcal{O}(m_t/\sqrt{\hat{s}})$ and explains the absence of $t$-channel behavior in \cref{eq:M-LL-+}. Due to a $t$-channel enhancement which is absent in other channels in the leading order, $W_L^{+}W_L^{-}\rightarrow t(+) \bar{t}(-)$ dominates top quark pair production from W-bosons and serve as the largest SM backgrounds in our analysis.

Furthermore, we numerically compute the amplitude of this channel by decomposing the top current into left-handed current and right-handed current parts separately. Explicitly, each $s$-channel diagram can be decomposed into the two subdiagrams with $\bar{t}_L \gamma^\mu t_L V_\mu$ and $\bar{t}_R \gamma^\mu t_R V_\mu$ with $V$ being either photon or the neutral $Z$ boson. The $t$-channel diagram only contains the left-handed current $\bar{t}_L \gamma^\mu b_L$ plus its conjugation $\bar{b}_L \gamma^\mu t_L$. 
It turns out that the summation of the left-handed current consisting of the $\bar{t}_L \gamma^\mu t_L$ part from $s$ channel diagram and the 
$t$ channel diagram dominates over the summation of the right-handed current part $\bar{t}_R \gamma^\mu t_R$ from the $s$ channel diagrams. This can be understood as follows. First, in the $t(+)\bar{t}(-)$ final helicity state, the Higgs diagram does not contribute, and the $t-$channel diagram only contains the left-handed current. In the high energy limit, the $s$-channel diagram exchanging $Z/\gamma$ bosons can be treated as restoring into $W_\mu^3$ field because the $W$ boson pairs can only interact with $W_\mu^3$ via the non-abelian self-interaction. While $W_\mu^3$ field can only couple to left-handed current. The right-handed current can be only from the finite $Z$ boson mass effect finally manifested by the $g'^2$ terms. Such behavior has a significant impact on the VLQ model, which will be discussed later on.

For $ZZ$-fusion channels, the partonic cross section is invariant under the CP transformation and the exchange of initial particle states. Consequently, only four distinct behaviors exist for all possible initial Z boson helicity combinations. Longitudinally polarized $Z$ boson states also dominate the partonic cross-section. The subdominant channels consist of one longitudinal polarized $Z$ boson plus the other transverse polarized $Z$ boson.
Both $t$-channel and $u$-channel diagrams are possible, resulting in both forward and backward enhancement of the differential distribution. This could also be understood because the top quark exchange is symmetric with respect to $\theta = \pi/2$. This contrasts with the $WW$-fusion case where the $t$-channel diagram exchanging the $b$ quark prefers the forward direction, and there is no $u$-channel diagram.  For production processes starting from transverse helicity modes, the cross-section from the opposite helicity initial state is larger than that of the same helicity.

For the diphoton-initiated processes, there are only two distinct combinations, with either the  same or opposite helicities of the photons.
One can see that the opposite helicity case (orange curve) is larger than the same-helicity case (red curve) in the high energy limit. Same-helicity process scales as $\hat{s}^{-1}$ asymptotically while the opposite-helicity process scales as $\hat{s}^{-1} \log \hat{s}$. A similar pattern also shows up in the $Z \gamma$ partonic cross section. It should be emphasized that the photon PDF from muon is around one order smaller than that from the electron at the same factorization scale, due to the two orders difference in the muon and electron mass. This results in decreased signal rate of diphoton initiated process at muon collider.

\subsection{Convolution with PDFs}

\begin{figure}[t!] 
  \centering
        \includegraphics[width=0.48\textwidth]{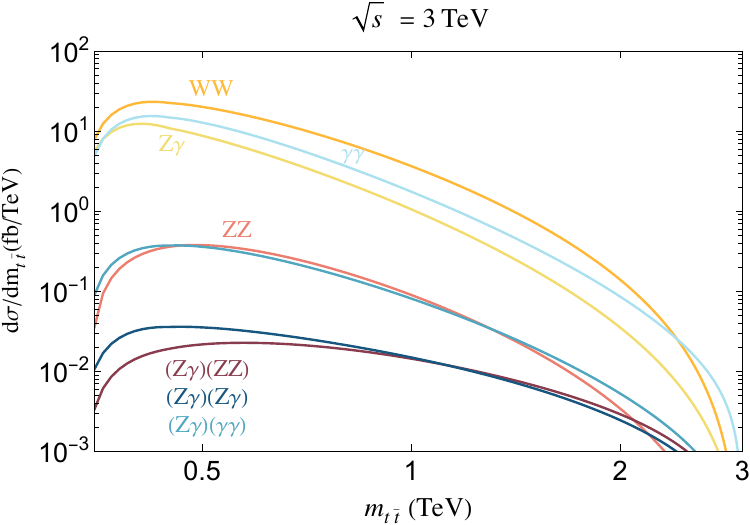}
        \includegraphics[width=0.48\textwidth]{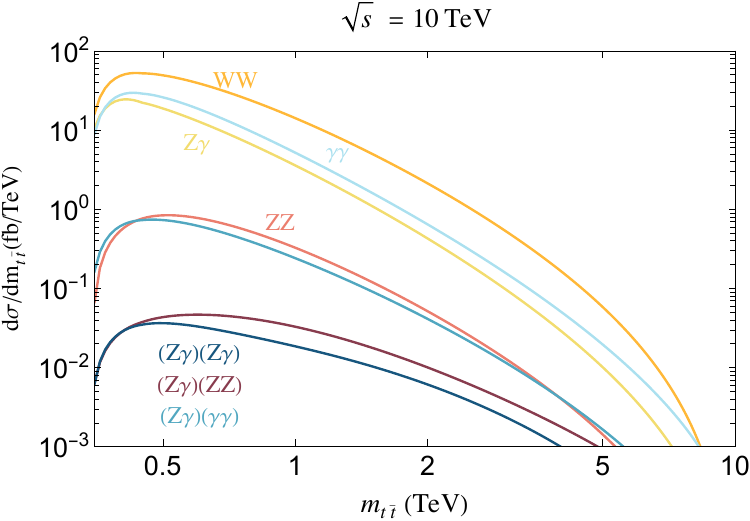}
\caption{The differential distribution for $\mu^+\mu^- \rightarrow t \bar{t} + X$ versus the invariant mass $m_{\bar{t}t}$ at a 3 TeV and a 10 TeV MuC.}

\label{fig:convolute_mtt_all_10TeV_3TeV}
\end{figure}

Having computed the partonic cross-section for all possible polarizations of the gauge boson, the next step is to convolute them with the electroweak gauge boson PDFs. We work with two benchmarks of the center of mass energies for the future muon collider: one is $\sqrt{s} = 3$ TeV with an integrated luminosity of 1 $\text{ab}^{-1}$, and the other one with $\sqrt{s} = $ 10 TeV with an integrated luminosity of 10 $\text{ab}^{-1}$. In this section, we first present the differential distribution with the invariant mass of the top quark pair in \cref{fig:convolute_mtt_all_10TeV_3TeV}.
The orders of $t\bar{t}$ cross-section contributions are $W W$, $\gamma \gamma$, $Z \gamma$ and $ZZ$ initial states. As we have seen in \cref{fig:partonic_distribution}, the partonic cross section of $W W$ is over two orders larger than that of $\gamma\gamma$ although the photon PDF is larger than $W$ boson. Due to the smallness of the $Z$ boson PDF, $ZZ$-fusion channel is suppressed while $Z \gamma$ sits in between. 
Beyond the simple PDF treatment, we also evaluate the importance of the interference effects between different electroweak gauge bosons. After all, the PDF treatment is a quasi-real approximation of the intermediate states, which are fully interfering. 
From the view of the full $2 \rightarrow 4$ process $\mu^+ \mu^- \rightarrow \mu^+ \mu^- t \bar{t}$, the intermediate off-shell gauge boson can be either $Z$ boson or the photon. One needs to first sum up the amplitudes of both processes before squaring them, namely
\begin{equation}
    \mathcal{M} = \sum_{i,j} \mathcal{M}_{V_{+,i} V_{-,j}} = \mathcal{M}_{Z Z} + \mathcal{M}_{\gamma \gamma} + \mathcal{M}_{Z \gamma}+ \mathcal{M}_{\gamma Z}
\end{equation}
where we denote $V_+$ as the neutral gauge boson radiated off $\mu^+$ and $V_-$ from $\mu^-$. After squaring, multiple interference terms should be handled carefully. Remembering that in computing the gauge boson PDF, there is the term $f_{Z\gamma}$ which accounts for the mixing from the splitting amplitude $\mathcal{M}^{\rm split}_{\mu\rightarrow \mu \gamma}$ and $\mathcal{M}^{\rm split}_{\mu\rightarrow \mu Z}$. Hence, the total cross section for $\mu^+ \mu^- \rightarrow \mu^+ \mu^- t \bar{t}$ should be 
\begin{equation}
    \sigma \sim \sum \int d x_1 d x_2 f_{i_- j_-}(x_1) f_{i_+ j_+}(x_2) \mathcal{M}_{i_- i_+} \mathcal{M}^*_{j_- j_+}
\end{equation}
For example, if $i_- = j_- = \gamma$,  $f_{i_- j_-}(x)$  equals to $f_\gamma(x)$. The case of $i_- \neq j_-$ corresponds to $f_{Z\gamma}$ function. We use the brackets to enclose $Z\gamma$ in \cref{fig:convolute_mtt_all_10TeV_3TeV} to show their contributions. It turns out that the interference terms are small compared to the diagonal terms where $i_{\pm}=j_{\pm}$.

\begin{figure}[!ht] 
  \centering
        \includegraphics[width=0.48\textwidth]{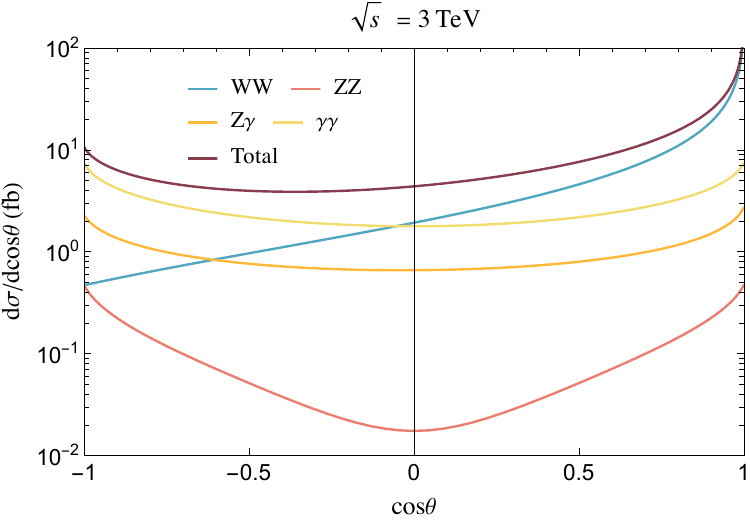}
         \includegraphics[width=0.48\textwidth]{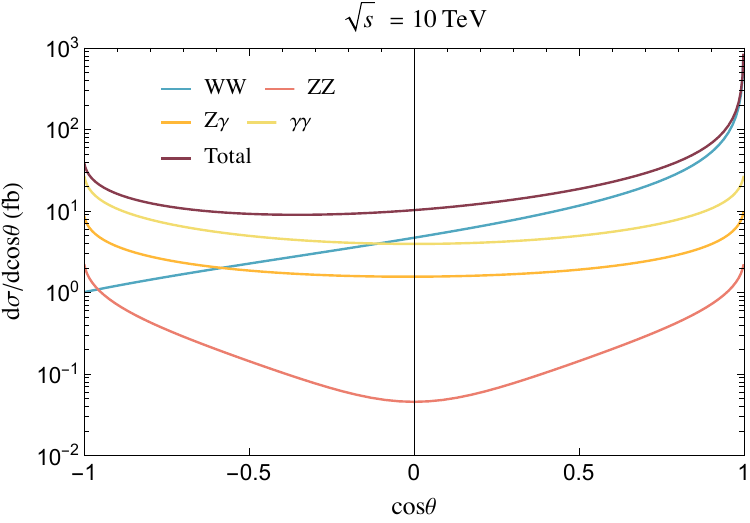}
    \caption{The differential angular distribution for $\mu^+\mu^- \rightarrow t \bar{t} + X$ at muon colliders with $\sqrt{s}$ = 3 TeV  and $\sqrt{s}$ = 10 TeV. }
   \label{fig:angular_3_10TeV}
\end{figure}

The angular distributions are shown in \cref{fig:angular_3_10TeV}. The left panel is for the 3 TeV MuC, and the right one is for 10 TeV MuC. The dominant channel $W^+W^-$ favors the forward direction due to the $t$-channel diagram exchanging intermediate $b$ quark. For the neutral gauge bosons channels, they are symmetric with respect to $\theta = \pi/2$.


\section{Projected Sensivities}\label{sec:result}

After exploring both the rate and the kinematics, we perform signal analysis and derive the sensitivity of the Top Yukawa coupling deviations for  3 TeV and 10 TeV muon colliders. This section presents the projected sensitivities for the anomalous Top Yukawa coupling and VLQ model.

\subsection{Top Yukawa Precision}\label{sec:ano_top_yukawa}

\begin{figure}[!ht] 
  \centering
               \includegraphics[width=0.492\textwidth]{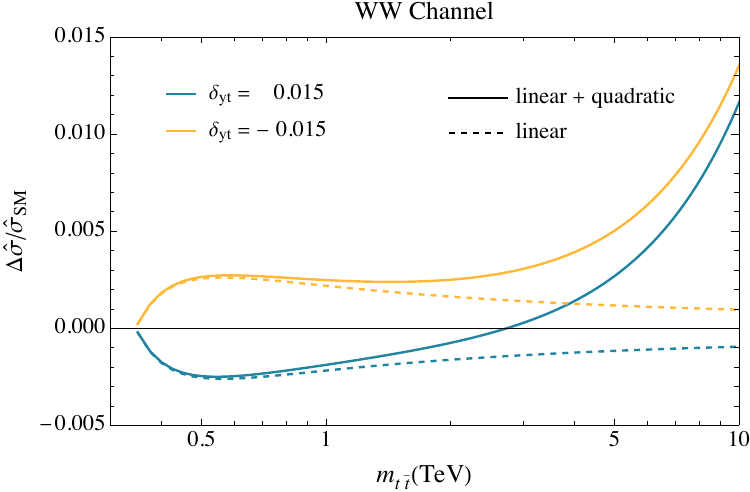}
         \centering
         \includegraphics[width=0.472\textwidth]{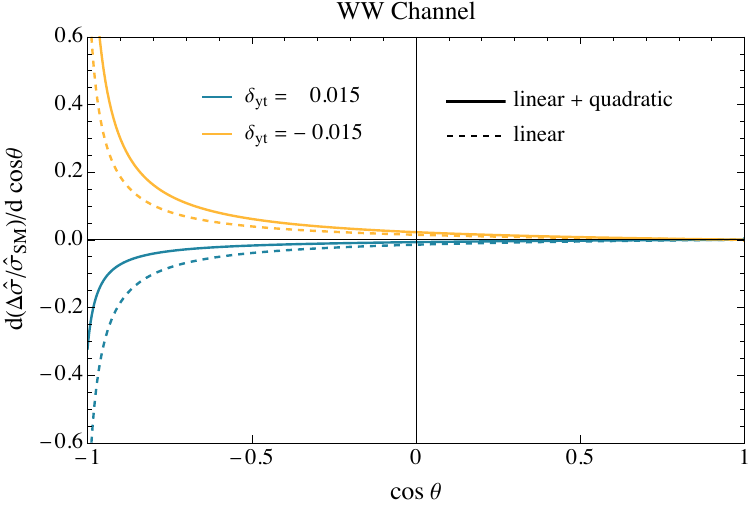}
    \caption{
    The deviation of $WW\to t\bar t$ cross-section from the Standard Model prediction value for benchmark choices of shift in Top Yukawa $\delta_{yt}$ as a function of (left panel:) partonic center of mass energy and (right panel:) scattering angle in the CM frame for a fixed partonic center of mass energy of 2~TeV. The solid and dashed lines are with and without the quadratic contributions, respectively.
    }
   \label{fig:dsigma_dyt_partonic}
\end{figure}

We show the relative change of the partonic cross section compared to the SM case due to the anomalous Top Yukawa coupling presented on the left panel of \cref{fig:dsigma_dyt_partonic}. Since the dominant signal comes from the WW channel after convoluting with PDF, so we will concentrate on this channel first to understand the leading contribution to the signal sensitivity. We can decompose the squared amplitude as
\begin{equation}
    |\mathcal{M}_{\rm tot}|^2 = |\mathcal{M}_{\rm SM} + \delta_{yt} \mathcal{M}_{\rm sig}|^2 = 
    |\mathcal{M}_{\rm SM}|^2 + 2 \delta_{yt} \,{\rm Re} \mathcal{M}_{\rm SM}^* \mathcal{M}_{\rm sig} + \delta_{yt}^2 |\mathcal{M}_{\rm sig}|^2,  
\end{equation}
here $\delta_{yt}$ is the shift in Top Yukawa, and $\mathcal{M}_{\rm sig}$ is the signal which will come from Higgs mediated diagram.

The dashed lines in \cref{fig:dsigma_dyt_partonic} show the signal-to-background ratio with linear in  $\delta_{yt}$ contributions (as a typical choice for EFT analysis),  and the solid lines include the terms proportional to $\delta_{yt}^2$. Here $\hat{\sigma}_{\rm SM}$ is the sum of all SM cross sections with all helicities. A constructive interference between the SM background and the anomalous signal for $\delta_{yt} < 0$ exists, and similarly, a destructive interference for $\delta_{yt}>0$ exists at linear order. As seen from the figure, if we only keep the linear term, while the modification to the $s$-channel Higgs amplitude scales as $\sqrt{\hat s}$ as shown in \cref{eq:non-unitary-growth}, $\Delta \hat{\sigma}/\hat{\sigma}_{\rm SM}$ approaches a constant asymptotically. The ratio grows linearly with $\hat{s}$ in the high energy limit if we also include the quadratic term. As we will show later, the sensitivity is dominated by the low invariant mass, linear regime. This implies the requirement for having systematic control at the sub-percent level. As an electroweak machine, it should be achievable, and we also want to emphasize here the importance of precision calculation. 
 
To understand the leading order behavior of these ratios,  one should first remember that only the cross section with final top quark helicities $(\pm,\pm)$ receives contributions from $\delta_{yt}$ in the process $W_L^+ W_L^- \rightarrow t \bar {t}$.
For the other two helicities, the Higgs exchange diagram is absent. More explicitly, we have
\begin{equation}
    \dfrac{\Delta \hat{\sigma}}{\hat{\sigma}_{\rm SM}} \sim \dfrac{\sum_{\lambda = \pm} 
   2 \delta_{yt} \text{Re} \mathcal{M}_{\lambda \lambda}^h \mathcal{M}^*_{\lambda \lambda} + \delta_{yt}^2|\mathcal{M}_{\lambda \lambda}^h|^2}
    {\sum |\mathcal{M}|^2}  \  .
\end{equation}
In the numerator, $\mathcal{M}_{\lambda \lambda}$ refers to all the SM amplitude with the top quark having the same helicities. $\mathcal{M}_{\lambda \lambda}^h$ denotes the amplitude for the Higgs-mediated diagram alone, which scales as $\sqrt{\hat s}$ in the high-energy limit.  
From \cref{tab:WWtt_cross_section}, we can infer that the amplitude of $W_L^+ W_L^- \rightarrow t(\pm) \bar{t}(\pm)$ scales like $1/\sqrt{\hat{s}}$ in the high energy regime  while the term proportional to $\delta_{yt}$ coming from the Higgs diagram scale as $\sqrt{\hat{s}}$. 
Thus the interference between these two terms will have no scaling behavior. 
On the other hand, the sum of the squared amplitudes in SM approaches constant at the leading order. 
Therefore, the ratio of the linear term in $\delta_{yt}$ of $\Delta \hat{\sigma}$ and $\hat{\sigma}_{\rm SM}$ becomes asymptotically flat since both have a leading order constant term. 
The term quadratic in $\delta_{yt}$ scales as $\hat{s}$ and so the ratio including both terms increases linearly with $\hat{s}$ asymptotically.  

\begin{figure}[!ht] 
	\centering 	\includegraphics[width=0.45\linewidth]{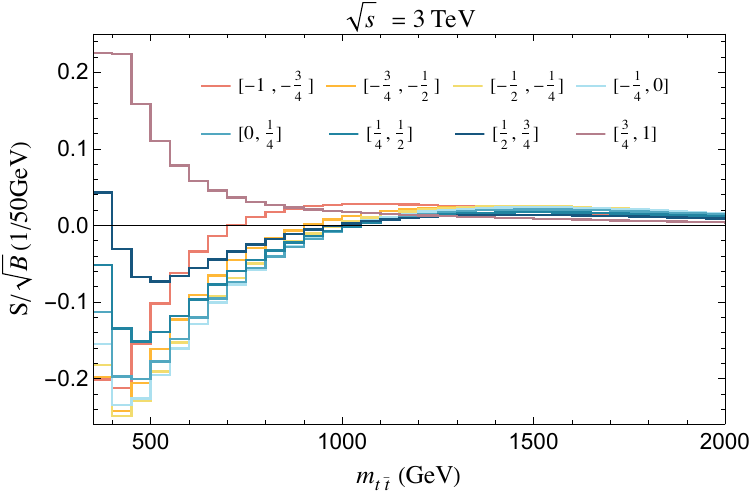}
\includegraphics[width=0.45\linewidth]{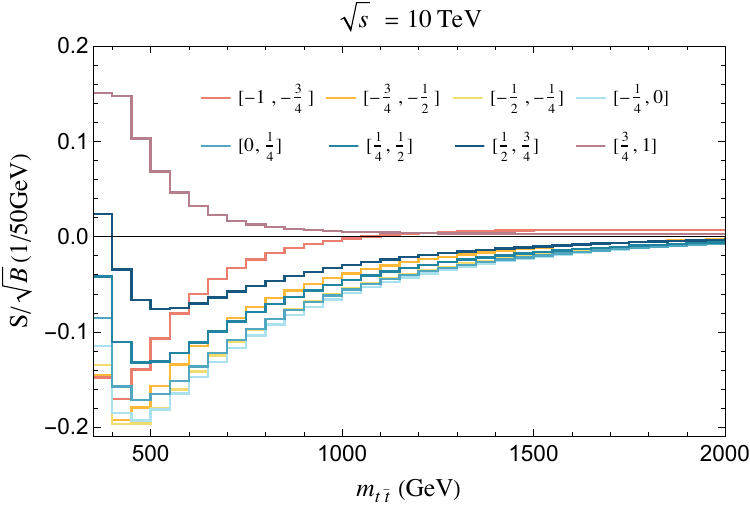}
	\caption{The significance of various bins in ranges of $m_{t\bar{t}}$ and $\cos \theta$ at 3 TeV and 10 TeV MuC. For illustration, we take $\delta_{yt}$ to be 0.015 for 10 TeV muon collider and 0.1 for 3 TeV muon collider, which is close to their one-sigma sensitivity. It can be seen from the figure that most of the sensitivity is coming from the interference term, near the threshold where the SM cross-section peaks and the quadratic term contribution is negligible. 
 }
	\label{fig:mtt-overlay-3TeV}
\end{figure}

The corresponding angular distribution of $\Delta \hat{\sigma}/\hat{\sigma}_{\rm SM}$ is shown on the right panel of \cref{fig:dsigma_dyt_partonic} for the partonic $\sqrt{\hat{s}} = 2$ TeV. Note that $\mathcal{M}_{\rm sig}$ has no angular dependence due to the spin 0 Higgs boson exchange. Therefore the angular dependence of the interference term is purely from the SM terms, and the term with quadratic $\delta_{yt}$ is flat. After dividing by the SM background contribution, the peak of the ratio occurs in the backward direction since the signal $\Delta \sigma$  is flatly distributed while the background favors the forward region.

To derive the sensitivity of the Top Yukawa coupling for the two muon collider benchmarks, we convolute the partonic signal cross sections with the PDF. For our analysis, we choose to bin the phase space in terms of the top quark pair invariant mass $ m_{t \bar{t}}$ and the outgoing polar angle in the CM frame with respect to the $\mu^+$ direction. Bin-by-bin signal significance after convolution is shown in \cref{fig:mtt-overlay-3TeV}, where the range of $\cos \theta$ is divided into eight equally sized bins, and $m_{t \bar{t}}$ is divided into 50 GeV bins to take into account finite resolution effects.  
From the figure, it is clear that our sensitivity will come from the linear term, which peaks near the threshold. 

\begin{figure}[!ht] 
  \centering
               \includegraphics[width=0.467\textwidth]{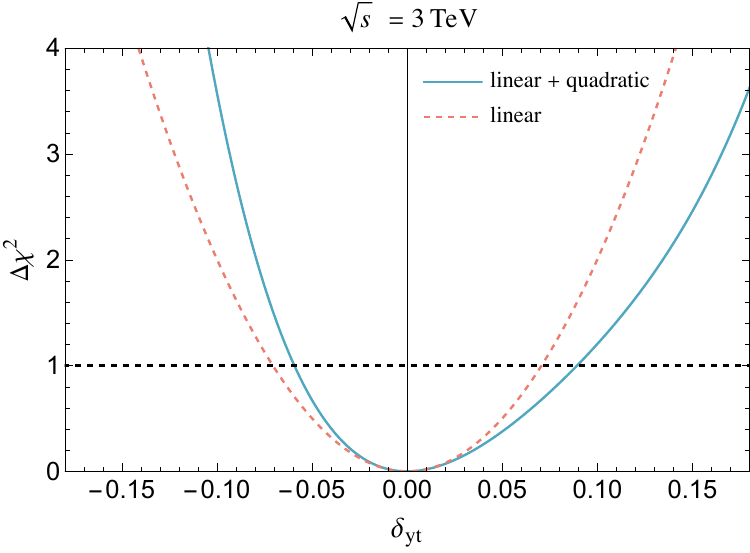}
         \centering
         \includegraphics[width=0.48\textwidth]{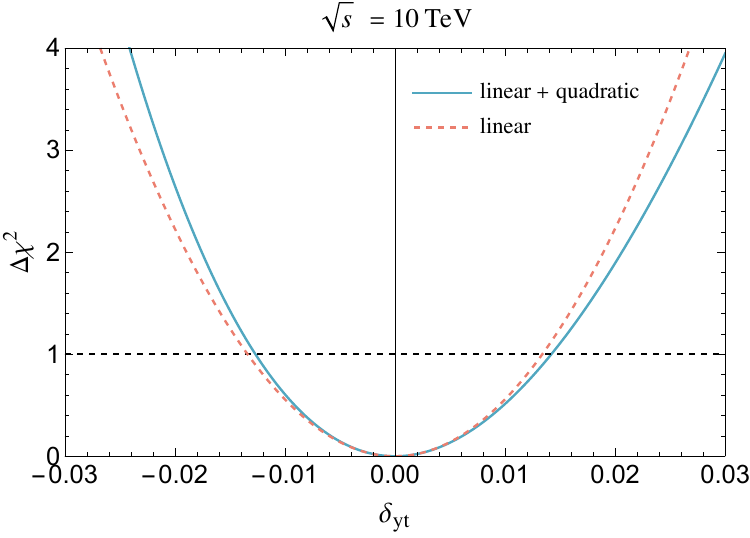}
\caption{The $\Delta \chi^2$ as a function of $\delta_{yt}$ after summation over bins for $\sqrt{s}$ = 3 TeV (left) and $\sqrt{s}$ = 10 TeV (right). Contributions from linear terms (interference piece) and quadratic terms are included in the $\Delta \chi^2$ analysis for the solid curves, and dashed curves only include the linear terms. For the solid curves, the $\Delta \chi^2$ asymmetry results from including the quadratic term.
}
\label{fig:dyt-chi-square}
\end{figure}

\begin{figure}[th]
    \centering
    \includegraphics[width=0.8\textwidth]{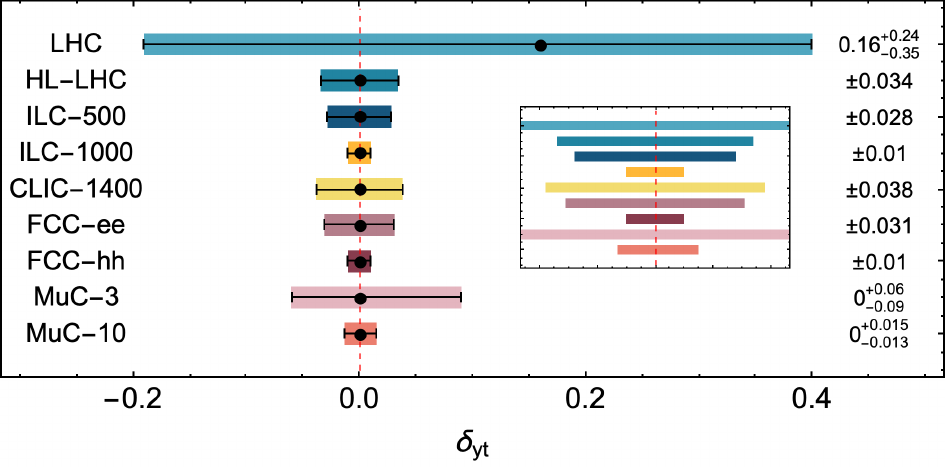}
    \caption{The reach plot for $\delta_{yt}$ at 1 $\sigma$ sensitivity on various colliders with their best projections through various processes. The bound on the Top Yukawa shift factor $\delta_{yt}$ from  (top to the bottom) the current LHC~\cite{CMS:2020djy,Negrini:2022gec}, future High Luminosity LHC~\cite{Cepeda:2019klc}, proposed ILC~\cite{Price:2014oca,ILCInternationalDevelopmentTeam:2022izu},CLIC~\cite{CLICdp:2018esa}, FCC-ee and FCC-hh~\cite{FCC:2018byv}. In the sub-panel, we show the zoom-in version to compare different bars in the smaller range.
    }
    \label{fig:dyt_compare}
\end{figure}

We perform a chi-square test to set bounds on anomalous Top Yukawa coupling. The results for the $\chi^2$ analysis are presented in \cref{fig:dyt-chi-square} for which  
we require the number of events in each bin to be at least 10. We compute the sensitivity $\Delta \chi_i^2 = S_i^2/B_i$ using the corresponding number of events for each bin. It is important to note that the signal ($S_i$) contains both the interference term and the quadratic term although the contribution of the quadratic term is negligible at the energy scale being considered. The total $\Delta \chi^2$ is obtained by summing over all the bins. This procedure is equivalent to a template double differential line-shape fit.

For a top quark pair, the branching ratios to various final states are $44\%$ (hadronic),  $44\%$ (semileptonic), $11\%$  (fully leptonic)\cite{ParticleDataGroup:2020ssz,ParticleDataGroup:2022pth}.\footnote{Here we split the $\tau$ final states into hadronic ones and leptonic ones with corresponding $\tau$ branch fractions.} In our final state considerations, we conservatively take only the fully hadronic and semi-leptonic $t\bar t$, as they are fully reconstructable without any sizable ambiguities. We also impose a cut on the outgoing $t$ angle $\theta$ with respect to $W^+$ in the partonic center of mass frame:
\begin{equation}
    10^{\circ}  \ < \ \theta \ < \ 170^{\circ},
\end{equation}
to ensure high reconstruction efficiency and avoid the forward regions where shielding is needed due to beam-induced backgrounds. These cuts do not notably change our sensitivity since most of the signal significance comes from the central bins in angular distribution, as seen in \cref{fig:ang-overlay-10TeV}.

The $\chi^2$ analysis presented in \cref{fig:dyt-chi-square} shows that the reach of $\delta_{yt}$ at one sigma C.L. is around $(-6 \%,9\%)$ for 3 TeV MuC and $(-1.3\%,1.4\%)$ for 10 TeV MuC. The corresponding 2 sigma C.L. exclusion are around $(-11\%,19\%)$ and $(-2.4\%,3.1\%)$ for 3 and 10 TeV MuC, respectively \footnote{If one use only the semi-leptonic $t\bar t$, the $95\%$ C.L. limit on the anomalous top coupling is projected to be  $5.6 \%$. This is highly consistent with the simulation-based study~\cite{Chen:2022yiu} at a 10 TeV MuC. }. It should be noted that in \cref{fig:dyt-chi-square}, we include the contributions from both the linear and quadratic terms of the signal. The resulting weight of these terms to the total $\Delta \chi^2$ is:
\begin{align}
\Delta \chi^2 \ &= \ 2.0 \times 10^2 \delta_{yt}^2 - 1.2 \times 10^3 \delta_{yt}^3 + 3.9 \times 10^3 \delta_{yt}^4 \ \ \ \text{(3TeV)} \label{eq:dchi23TeV}\\ 
\Delta \chi^2 \ &= \ 5.6 \times 10^3 \delta_{yt}^2 - 4.8 \times 10^4 \delta_{yt}^3 + 2.6 \times 10^5 \delta_{yt}^4 \ \ \ \text{(10TeV)} 
\label{eq:dchi210TeV}
\end{align}
As can be seen, the inclusion of quadratic term results in producing asymmetric $\Delta \chi^2$ distribution. Furthermore, at one sigma C.L for both these colliders, if we neglect the quadratic term,  $\delta_{yt}$ reach would be 7\% for 3 TeV and 1.33\% for 10 TeV case. \cref{fig:dyt-chi-square} also shows the results for only including the interference effects in dashed lines, which is conventionally done for EFT analysis. We can see the span of the sensitivity is very similar to the results, including the quadratic terms. One can obtain the corresponding $\Delta \chi^2$ expressions by dropping the last two terms in the above equations.

For more general Higgs precision fits, the relative contributions from the $WW$-fusion and $ZZ$-fusion channels need to be specified. Here, we also write down the $\Delta\chi^2$ in the widely used $\kappa$-framework~\cite{Han:2013kya,Asner:2013psa,CEPC-SPPCStudyGroup:2015csa,deBlas:2019rxi,deBlas:2022aow}. Defining $\kappa_W = g_{WWh}/g^{\rm SM}_{WWh}$, $\kappa_Z = g_{ZZh}/g^{\rm SM}_{ZZh}$, $\kappa_t = y_t/y_t^{\rm SM}$ in the kappa-framework, we further perform the chi-square analysis for 3 TeV collider:
\begin{equation}
\begin{split}
\Delta \chi^2 \ &= \ 2.2 \times 10^2 (\kappa_W \kappa_t-1)^2 + 2.9 (\kappa_Z \kappa_t-1)^2 - 23 (\kappa_W \kappa_t-1)(\kappa_Z \kappa_t-1) \\
&-1.2 \times 10^3 (\kappa_W \kappa_t-1)^3 
 + 10 (\kappa_Z \kappa_t-1)^3+92(\kappa_W \kappa_t-1)^2(\kappa_Z \kappa_t-1) \\
 &-1.3 \times 10^2(\kappa_W \kappa_t-1)(\kappa_Z \kappa_t-1)^2 +3.1\times 10^3(\kappa_W \kappa_t-1)^4+ 38(\kappa_Z \kappa_t-1)^4 \\
&+6.9\times 10^2(\kappa_W \kappa_t-1)^2 (\kappa_Z \kappa_t-1)^2 
\end{split}
\label{eq:kappadchi23TeV}
\end{equation}
and for 10 TeV:
\begin{equation}
\begin{split}
\Delta \chi^2 \ &= 6.2 \times 10^3 (\kappa_W \kappa_t-1)^2 + 1 \times 10^2  (\kappa_Z \kappa_t-1)^2 -  6.6 \times 10^2  (\kappa_W \kappa_t-1)(\kappa_Z \kappa_t-1) \\
& -4.7\times 10^4(\kappa_W \kappa_t-1)^3 
 + 4.4 \times 10^2 (\kappa_Z \kappa_t-1)^3+4.0 \times 10^3(\kappa_W \kappa_t-1)^2(\kappa_Z \kappa_t-1) \\
 &- 5.1 \times 10^3(\kappa_W \kappa_t-1)(\kappa_Z \kappa_t-1)^2 +2.1 \times 10^5(\kappa_W \kappa_t-1)^4+ 2.6 \times 10^3(\kappa_Z \kappa_t-1)^4 \\
&+4.7 \times 10^4(\kappa_W \kappa_t-1)^2 (\kappa_Z \kappa_t-1)^2 
\end{split}
\label{eq:kappadchi210TeV}
\end{equation}
Similarly, one can obtain the interference alone contribution by dropping all terms other than the first lines of the above equations. These equations also reveal that the sensitivity reach is dominated by the $WW$ channel. Note that here, we assume no forward muon tagging, which means the $ZZ$-fusion and $WW$-fusion are indistinguishable at the analysis level. In other words, they share the same backgrounds. If one can tag the forward muons and hence reduce the $ZZ$-fusion backgrounds, the relative importance and contribution will change. 

The $\delta_{yt}$ reach at one sigma C.L. is be (-6\%,8\%) for 3TeV MuC and (-1.2\%,1.4\%) for 10TeV MuC. In \cref{fig:dyt_compare}, we show the one sigma sensitivity on $\delta_{yt}$ at the current LHC, HL-LHC, ILC-500, ILC-100, CLIC, FCC-ee, and FCC-hh, and compare them with our results at muon colliders of 3 and 10 TeV. At present, the LHC experimental constraint is at the $O(10\%)$ level. The future 100 TeV FCC-hh can probe the Top Yukawa coupling to around 1\%.  We show here the muon collider can achieve comparable sensitivity.

\subsection{VLQ Model}

\begin{figure}[!ht] 
  \centering
               \includegraphics[width=0.48\textwidth]{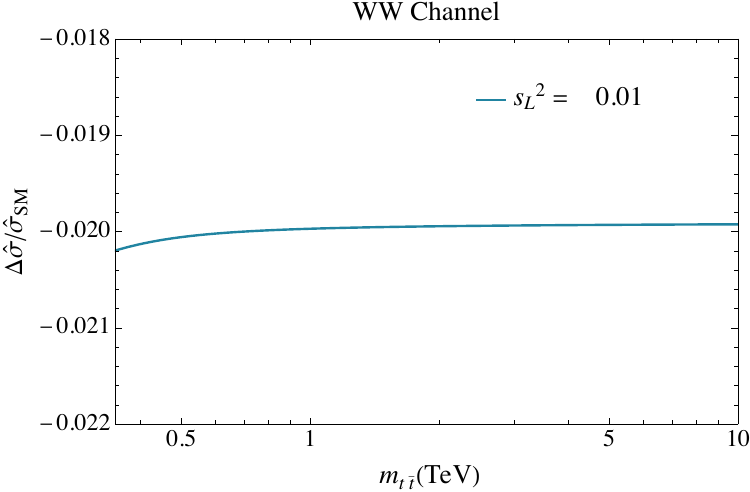}
         \includegraphics[width=0.47\textwidth]{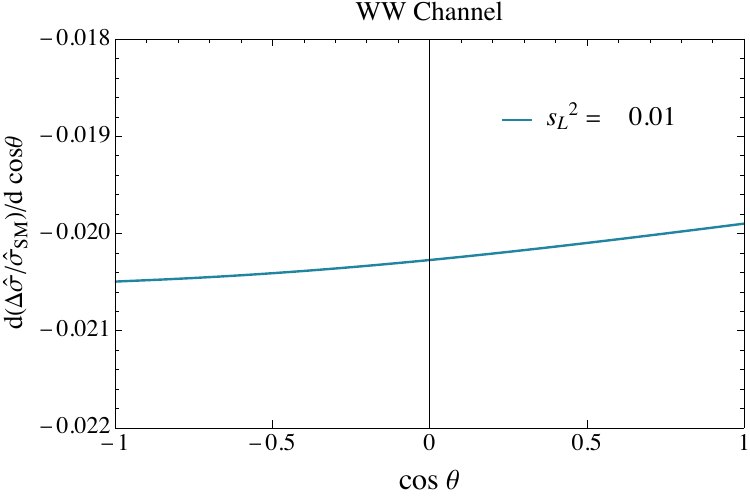}
    \caption{ The relative shift of the partonic cross section for the WW channel in VLQ model as a function of $\sqrt{\hat{s}}$ and the polar angle $\cos\theta$.}
   \label{fig:VLQ_WW_partonic}
\end{figure}

Now we study this channel in the VLQ model, where Top Yukawa modification correlate with other coupling modifications. 
Analogous to our previous discussion, we first focus on the deviation in the partonic cross-section of the $W^+W^- \rightarrow t \bar{t}$ channel which is shown in \cref{fig:VLQ_WW_partonic} for $s_L^2 = 0.01$. Since this is a UV-complete model, we keep all the higher order terms of $\delta_{yt}$ or $s_L^2$. We observe that the relative change decreases when the center-of-mass energy increases from the thresh-hold energy and becomes almost flat as $\sqrt{\hat{s}}$ exceeds around 1 TeV. To understand this result, first, we note that the dominant (helicity) amplitude for the SM background is asymptotically flat. Since we are working in the weakly-coupled UV-completed model, the amplitude should also be unitarized. This explains why the two distributions are relatively flat. It is interesting to note that the angular distribution of the BSM-to-SM ratio $\Delta \hat{\sigma} / \hat{\sigma}_{\rm SM}$ is also nearly flat, which suggests that the signal follows the shape of the background. In other words, the VLQ signal shifts the SM cross-section. 

The shifting can be explained by noting that the leading order contribution for $W^+_L W^-_L \rightarrow t \bar{t}$ is dominated by the left-handed current as we have emphasized in \cref{subsec:partonic_cross}.  For our VLQ analysis, the $t_L$ field is rescaled by $c_L$. Hence the amplitudes of the two channels are shifted by a  common factor of $-s_L^2$. This explains why the relative shifting factor for the cross section is approaching $-2 s_L^2$ in the high energy regime and is almost independent of the theta angle. 

On the other hand, we can also compute using the Goldstone boson equivalence theorem. As discussed in \cref{sec:theoretical_frame}, if there is only a single operator EFT $\mathcal{O}_y^t$, a new contact vertex $\phi^+\phi^- t \bar{t}$ is generated, leading to the energy growing behavior. While in the VLQ case,  two other operators can generate the new coupling $i\left(\partial_\mu \phi^{+} \bar{t}_L \gamma^\mu b_L-\partial_\mu \phi^{-} \bar{b}_L \gamma^\mu t_L\right)$ that modify the $t$-channel diagram amplitude which also has the energy growing behavior. The correlated Wilson coefficients guarantee the cancellation of individual unitarity violations.

\begin{figure}[!ht] 
  \centering
        \includegraphics[width=0.48\textwidth]{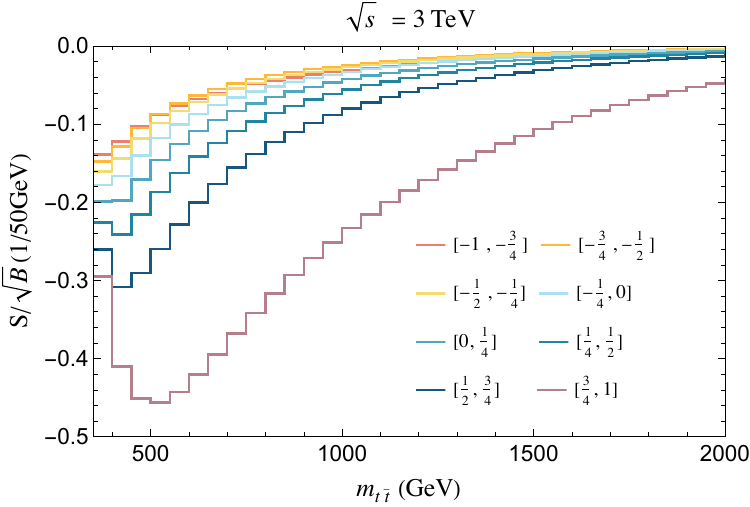}
         \includegraphics[width=0.48\textwidth]{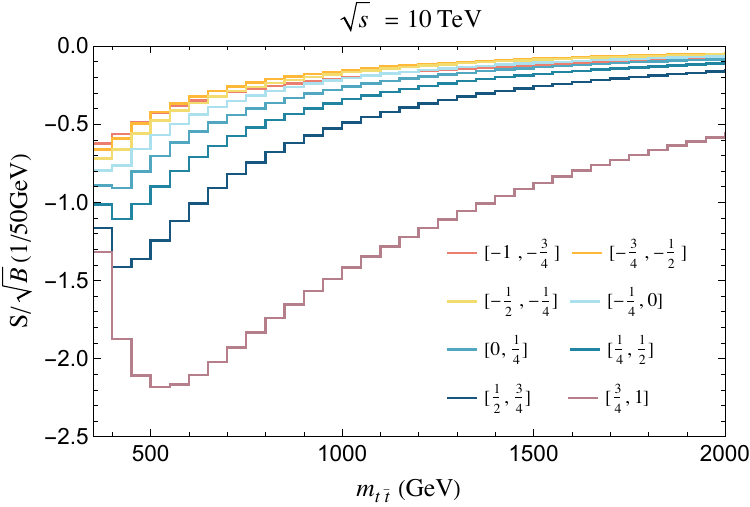}
    \caption{ $S/\sqrt{B}$ for $s_L^2$ = 0.01 for both a 3 TeV (left) and a 10 TeV (right) MuC. We cut the whole space in terms of the invariant mass $m_{t\bar{t}}$ and the angular variable $\cos\theta$. Different colors denote the bin range of $\cos\theta$. }
   \label{fig:VLQ_S_sqrtB}
\end{figure}

\begin{figure}[!ht] 
  \centering
        \includegraphics[width=0.48\textwidth]{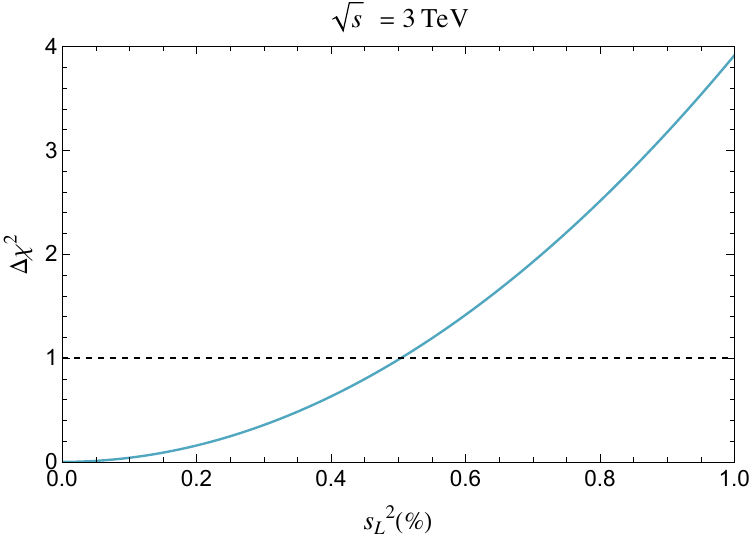}
         \includegraphics[width=0.48\textwidth]{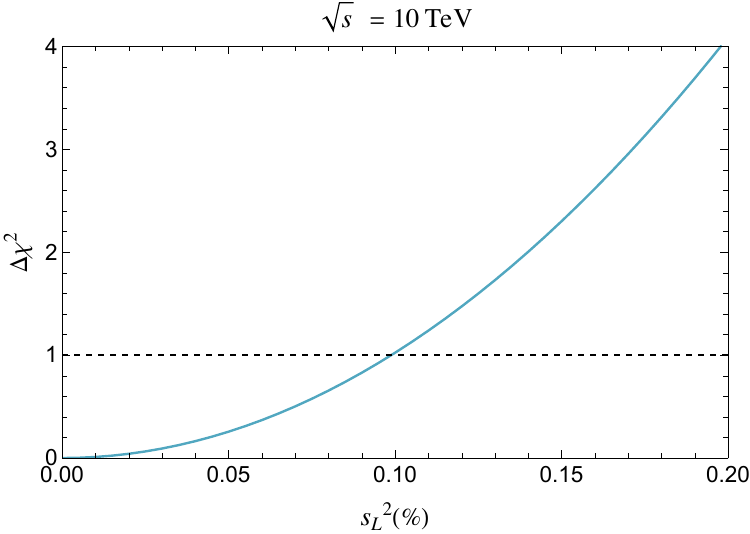}
\caption{The $\Delta \chi^2$ as a function of $s_L^2 = -\delta_{yt}$  after summation over bins for $\sqrt{s}$ = 3 TeV (left) and $\sqrt{s}$ = 10 TeV (right) in VLQ model.  }
\label{fig:chi_VLQ_sL2}
\end{figure}


In \cref{fig:VLQ_S_sqrtB}, we show the $S/\sqrt{B}$ distribution over the top quark pair invariant mass 
$m_{t\bar{t}}$ after dividing the phase space into various bins. Curves with different colors refer to the angular bins. The peak values for the significance occur at low $m_{t\bar{t}}$ regime as well as the forward region.  As $S/B$ is almost a constant,  we get stronger sensitivity in the forward direction and low $m_{t\bar{t}}$ region where the statistics are large.


\begin{figure}[th]
    \centering
    \includegraphics{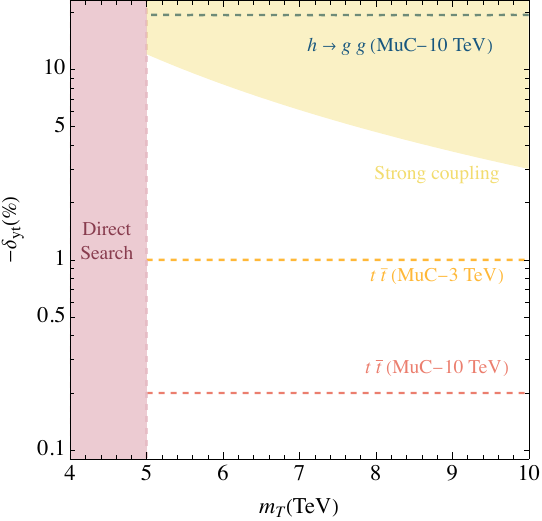}
    \caption{The 95\% C.L. exclusion sensitivity for the parameters in VLQ model. The blue dashed line is from the top quark pair production and the Higgs decaying to di-gluon measurement~\cite{Forslund:2022xjq}. The orange and pink dashed lines refer to the exclusion sensitivity at 95\% exclusion limit. The light-yellow region denotes the strong coupling regime where the UV Yukawa coupling $\lambda_0$ is larger than $O(10)$. We expect the region for $m_T < 5$ TeV can be covered by the direct search, as the $T \bar{T}$ pair production is kinematically allowed. 
    }
    \label{fig:VLQ_para2}
\end{figure}

We follow the same procedure described in the previous section to estimate the projected sensitivity and show the results in \cref{fig:chi_VLQ_sL2}. The anomalous Top Yukawa coupling shift can be tested to around 0.5\% at 3 TeV MuC and 0.1\% at 10 TeV MuC at one sigma level. In \cref{fig:VLQ_para2}, we show various 95\% exclusion regions on the  $m_T-\delta_{yt}$ plane. The blue, orange, and pink dashed lines refer to the bound from the Higgs decaying to di-gluon precision measurement at 10 TeV MuC, the $t\bar{t}$ channel at 3 TeV MuC and 10 TeV MuC respectively. The coupling is strong in the light-yellow region, violating perturbative unitarity in the Higgs-mediated four-fermi scattering process, with the UV Yuakawa coupling $ \lambda_0 > O(10)$ in \cref{eq:VLQ_dyt_exp}. The shaded region for $m_T < 5$ TeV is denoted as ``Direct Search'', where the $T\bar{T}$ production is kinematically allowed. We also made an estimate of the reach of the single heavy VLQ search at the 10 TeV MuC, and this channel can exclude $\delta_{yt}$ to be around one percent level. However, given the sensitivity dependence on the detailed realization of VLQ, e.g., the existence of heavy bottoms, we do not show the estimation here. For a similar reason, we also do not show electroweak precision from future Z-pole programs in this plot.


\section{Conclusion}\label{sec:conclusion}

In this work, we have examined the prospects for measuring the Top Yukawa coupling at upcoming high-energy muon colliders, focusing on two benchmark center-of-mass energies: 3 TeV and 10 TeV. We thoroughly investigated the partonic cross sections of the processes $V V \rightarrow t \bar{t}$ for various helicities and delved into the differential kinematical distributions. Upon convolution with the electroweak gauge boson PDFs, it becomes evident that the dominant partonic channel is the scattering of longitudinal $W$ bosons into a right-handed top quark and a left-handed anti-top-quark.

We utilized two scenarios to characterize deviations from the Standard Model (SM) Top Yukawa coupling. We first considered the case in which a single dimension-six EFT operator can encapsulate the influences of new physics, denoted as $\mathcal{O}_{y}^t$. Here, the primary effect is an anomalous deviation in the Top Yukawa coupling. Consequently, the amplitude of $W^+ W^- \rightarrow t \bar{t}$ rises with the partonic center of mass energy, leading to significant discrepancies from SM predictions at larger $m_{t\bar{t}}$. In our second approach, taking inspiration from a UV-completed model, we introduced a pair of singlet vector-like quarks. Post electroweak symmetry breaking, a mixture occurs between the top quark and the heavy VLQ, altering the Top Yukawa coupling. In this scenario, the amplitude or cross-section's relative shift remains mostly flat across a broad range of  $m_{t\bar{t}}$.

Constraints on the Top Yukawa deviation can be established, assuming future measurements agree with SM. Segmenting the entire phase space according to the production polar angle and the invariant mass of the top quark pair allows for chi-square analysis. For the EFT model, the Top Yukawa measurement precision approaches the percent level, providing valuable measurements of Top Yukawa beyond HL-LHC. The results are summarized in \cref{fig:dyt_compare}. Furthermore, we provide chi-square data for global fit to the Higgs signal, in \cref{eq:dchi23TeV,eq:dchi210TeV,eq:kappadchi23TeV,eq:kappadchi210TeV}. Importantly, compared to the conventional channel of $t\bar{t}h$~\cite{Forslund:2022xjq,Forslund:2023reu}, our method is more than one order of magnitude better in projected precision and also sensitive to the sign of the Top Yukawa deviation.  Conversely, in the VLQ model, this precision is further improved to nearing  per mil level. These high-precision measurement also requires precision calculation from the theory side to ensure their robustness. We estimated the leading order perturbative correction would be at \%-level. \\

\textit{Acknowledgements:}
We want to thank K. Hagiwara, T. Han, Y. Ma, P. Meade and A. Wulzer for their helpful discussions, and J. Fan and M. Reece for participation in the earlier part of this study.
K.F.L., Z.L., and I.M. are supported in part by the DOE grant DE-SC0022345 and  DOE grant DE-SC0011842. 
Z.L., K.F.L., and LTW acknowledge the Aspen Center for Physics where the final part of this work is completed, which is supported by National Science Foundation grant PHY-2210452. The work of L.T.W. is supported by the DOE grant DE-SC-0013642.
The data associated with the figures in this paper can be accessed via \href{https://github.com/ZhenLiuPhys/WWtt-MuC}{GitHub}.

\clearpage
\begin{appendix}
\section{Helicity Amplitude} \label{sec:helicity amplitudes}
In this section, we focus on the amplitude of $W^+ W^- \rightarrow t \bar{t}$ and its analytic expressions.
 The total amplitude for a given helicity configuration can be written as the sum of contributions from four channels:
\begin{equation}
    \mathcal{M}_{h_{W^+} h_{W^{-}}, h_t h_{\bar{t}}} \ = \ \mathcal{M}^{\gamma} + \mathcal{M}^{Z} + \mathcal{M}^{b} + \mathcal{M}^h  
\end{equation}
where the $h$ in the subscript refers to the corresponding helicities. $\mathcal{M}^{\gamma/Z/h}$ refer to the $s$-channel contribution and $\mathcal{M}^b$ refer to the $t$-channel diagram. We will compute the amplitude in the massless $b$-quark limit. Also, we choose the incoming $W^+$ to be aligned along the z-axis and define the scattering angle $\theta$ as the polar angle between $W^+$ and the $t$-quark in the partonic center-of-mass frame. Thus, with this definition, we have in the massless $b$ quark limit:
\begin{equation}
    s  =  4 E_t^2  = 4 E_W^2 \ \ , \ \ t \ = - \frac{s}{4} (\beta_t^2 + \beta_W^2 - 2 \beta_t \beta_W \cos \theta)
\end{equation}
Here, $E_{t,W}$ is the center of mass energy of each particle, and $\beta_{t,W}^2 = 1 - 4m_{t,W}^2/s$. In this section, all the variables are partonic variables, and we will drop the hat for abbreviation as there is no ambiguity. 
For our calculations, we use the explicit form of four-spinors and polarization vectors in the center of mass frame and calculate the helicity amplitudes. Wigner d-functions are  explicitly given by:
\begin{align}
 &  d_{0,0}^2 \ = \frac{1}{2} \left(3 \cos^2\theta -1 \right), ~~~ d_{2,0}^2 \ = \sqrt{\frac{3}{8}} \sin^2\theta \\
& d_{1,0}^2 \ = - d_{-1,0}^2 \ = \ - \sqrt{\frac{3}{8}} \sin 2\theta \\
 & d_{2,1}^2 \ = - d_{-2,-1}^2 \ = \ - \frac{1}{2} \left(1 + \cos \theta \right) \sin \theta \\ 
&   d_{2,-1}^2 \ = - d_{-2,1}^2 \ = \ - \frac{1}{2} \left(1 - \cos \theta \right) \sin \theta \\
 & d_{1,1}^2 \ =  d_{-1,-1}^2 \ = \ \frac{1}{2} \left( 2 \cos^2 \theta +  \cos \theta - 1 \right) \\ 
&   d_{1,-1}^2 \ =  d_{-1,1}^2 \ = \ \frac{1}{2} \left( - 2 \cos^2 \theta +  \cos \theta + 1 \right) \\
&   d_{1,1}^1 \ =  d_{-1,-1}^1 \ = \ \frac{1}{2} \left(1 + \cos \theta \right) \\ 
&  d_{1,-1}^1 \ =  d_{-1,1}^1 \ = \  \frac{1}{2} \left(1 - \cos \theta \right)  \\
&  d_{1,0}^1 \ = - d_{-1,0}^1 \ = \ - \sqrt{\frac{1}{2}} \sin \theta  \\
&  d_{0,0}^1 \ = \cos \theta, ~~~d_{0,0}^{0}\ = 1 \\
& d_{m,m'}^j \ = (-1)^{m-m'}d^j_{m',m} \ = \ d_{-m',-m}^j\ 
\end{align}

\begin{itemize}
\item $(00,++)$
\begin{align}
    \mathcal{M}_{00,++}^{\gamma} \ &= \  \frac{2}{3} \ \left[2 \sqrt{2}G_F m_t s_w^2   s^{1/2}  \beta_W \left(3 -\beta_W^2 \right) \right] d_{0,0}^1 
    \label{eq:a14}
\\
    \mathcal{M}_{00,++}^{Z} \ &= \  \frac{2}{3} \ \left[\frac{1}{ \sqrt{2}}G_F m_t    \left(\frac{3}{2} - 4s_w^2 \right)  \frac{s^{3/2}}{s-M_Z^2}  \beta_W \left(3 -\beta_W^2 \right) \right] d_{0,0}^1 
       \label{eq:a15}
\\
    \mathcal{M}_{00,++}^{b} \ &= \  \frac{1}{2 \sqrt{2}} G_F m_t \frac{ s^{3/2}}{t}   \left[ \beta_t \left( \frac{1}{3} + \beta_W^2 \right) +  \beta_W \left( 1 - \beta_W^2\right)  d^1_{0,0} - \frac{4}{3} \beta_t d^2_{0,0}  \right]
       \label{eq:a16}
  \\
\mathcal{M}_{00,++}^{h} \ &= \  \frac{1}{{2}} G_F y_t v \frac{s^{3/2}}{s - m_H^2} \beta_t \left(1 + \beta_W^2\right)
   \label{eq:a17}
\end{align}

\item$(00,+-)$
\begin{align}
    \mathcal{M}_{00,+-}^{\gamma} \ &= \ - \frac{2}{3} \ \left[2 G_Fs_w^2  s \beta_W 
 \left(3 -\beta_W^2 \right) \right] d_{0,1}^1 
    \\
    \mathcal{M}_{00,+-}^{Z} \ &= \ \ - \frac{2}{3} \ \left[\frac{1}{2} G_F  \left(\frac{3}{2}\left(1 - \beta_t \right) - 4s_w^2 \right) \frac{s^2}{s-M_Z^2} \beta_W  \left(3 -\beta_W^2 \right) \right] d_{0,1}^1 
    \\
    \mathcal{M}_{00,+-}^{b} \ & = \   \frac{1}{2}G_F  \frac{s^2}{t} (1 - \beta_t)  \left[ \left( \beta_W \beta_t - \frac{1}{2} \beta_W \left( 1 - \beta_W^2 \right)\right) d^1_{0,1}  + \sqrt{\frac{2}{3}} \beta_t d^2_{0,1}  \right]
    \\
\mathcal{M}_{00,+-}^{h} \ &= \ 0
\end{align}

\item$(00,-+)$
\begin{align}
    \mathcal{M}_{00,-+}^{\gamma} \ & = \ \frac{2}{3} \ \left[2 G_Fs_w^2   s \beta_W \left(3 -\beta_W^2 \right) \right] d_{0,-1}^1 
    \label{Ph1}
    \\
    \mathcal{M}_{00,-+}^{Z} \ &= \  \frac{2}{3} \ \left[ \frac{1}{2} G_F\left(\frac{3}{2}\left(1 + \beta_t \right) - 4s_w^2 \right)  \frac{s^2}{s-M_Z^2} \beta_W  \left(3 -\beta_W^2 \right)\right] d_{0,-1}^1 
    \\
    \mathcal{M}_{00,-+}^{b} \ & = - \  \frac{1}{2}G_F  \frac{s^2}{t} \left(1+\beta_t \right)  \left[ \left( -\beta_W \beta_t - \frac{1}{2} \beta_W \left( 1 - \beta_W^2 \right)\right) d^1_{0,-1}  + \sqrt{\frac{2}{3}} \beta_t d^2_{0,-1}  \right]
    \\
\mathcal{M}_{00,-+}^{h} \ & = \ 0
\end{align}

\item $(00,--)$
\begin{align}
    \mathcal{M}_{00,--}^{\gamma} \ & = \ - \frac{2}{3} \ \left[2 \sqrt{2}G_F m_t s_w^2  s^{1/2} \beta_W \left(3 -\beta_W^2 \right) \right] d_{0,0}^1
      \label{eq:a26}
    \\
    \mathcal{M}_{00,--}^{Z} \ & = \ - \frac{2}{3} \ \left[\frac{1}{ \sqrt{2}}G_F m_t   \left(\frac{3}{2} - 4s_w^2\right)  \frac{s^{3/2}}{s-M_Z^2}   \beta_W  \left(3 -\beta_W^2 \right) \right] d_{0,0}^1 
      \label{eq:a27}
    \\
    \mathcal{M}_{00,--}^{b} \ & = \ -  \frac{1}{2 \sqrt{2}} G_F m_t \frac{ s^{3/2}}{t}   \left[  \beta_t \left(  \frac{1}{3} + \beta_W^2 \right) +  \beta_W \left( 1 - \beta_W^2\right)  d^1_{0,0} - \frac{4}{3} \beta_t d^2_{0,0}  \right]
          \label{eq:a28}
    \\
\mathcal{M}_{00,--}^{h} \ & = \ - \frac{1}{{2}} G_F y_t v \frac{s^{3/2}}{s - m_H^2} \beta_t \left(1 + \beta_W^2\right)
  \label{eq:a29}
\end{align}

\end{itemize}

\begin{itemize}
    \item $(-0,++)$
\begin{align}
    \mathcal{M}_{-0,++}^{\gamma} \ & = \  \frac{2}{3} \ \left[8 \sqrt{2} G_F m_t m_W s_w^2  \beta_W     \right] d_{-1,0}^1 
       \\
    \mathcal{M}_{-0,++}^{Z} \ & = \ \frac{2}{3} \ \left[ 2 \sqrt{2} G_F m_t m_W \left(\frac{3}{2} - 4s_w^2 \right) \frac{s}{s - M_Z^2}    \beta_W    \right] d_{-1,0}^1 
    \\
    \mathcal{M}_{-0,++}^{b} \ & = \  - \frac{1}{ \sqrt{2}} G_F m_t m_W   \frac{s}{t}   \left[ \left( - \beta_t \beta_W - \beta_t - \beta_W \left(1 - \beta_W \right) \right) d_{-1,0}^1  +\frac{2 }{\sqrt{3}} \beta_t  d_{-1,0}^2  \right] \\
    \mathcal{M}_{-0,++}^{h} \ & = \ 0
\end{align}

    \item $(-0,+-)$
\begin{align}
    \mathcal{M}_{-0,+-}^{\gamma} \ & = \  - \frac{2}{3} \  \left[8 G_Fm_W s_w^2 s^{1/2} \beta_W  \right] d_{-1,1}^1 
       \\
    \mathcal{M}_{-0,+-}^{Z} \ & = \ - \frac{2}{3} \ \left[ 2 G_Fm_W\left(\frac{3}{2} \left(1 - \beta_t \right) - 4s_w^2\right) \frac{s^{3/2}}{s - M_Z^2} \beta_W   \right] d_{-1,1}^1 
    \\
    \mathcal{M}_{-0,+-}^{b} \ & = \   \frac{1}{2} G_F m_W   \frac{s^{3/2}}{t} (1 -  \beta_t) \left[  \left(- \beta_W \left(1- \beta_W \right) + \beta_t \beta_W \right) d^1_{-1,1}  +\beta_t d^2_{-1,1}   \right]
    \\
    \mathcal{M}_{-0,+-}^{h} \ & = \ 0
\end{align}
    \item $(-0,-+)$
\begin{align}
    \mathcal{M}_{-0,-+}^{\gamma} \ & = \    \frac{2}{3} \  \left[8 G_Fs_w^2m_W s^{1/2} \beta_W \right] d_{-1,-1}^1 
       \\
    \mathcal{M}_{-0,-+}^{Z} \ & = \  \frac{2}{3} \ \left[ 2 G_Fm_W \left(\frac{3}{2} \left(1 + \beta_t \right) - 4s_w^2 \right) \frac{s^{3/2}}{s - M_Z^2} \beta_W \right] d_{-1,-1}^1 
    \\
    \mathcal{M}_{-0,-+}^{b} \ & = \ - \frac{1}{2}  G_F m_W  \frac{s^{3/2}}{t} (1+ \beta_t) \left[  \left(- \beta_W \left(1- \beta_W \right) - \beta_t \beta_W \right) d^1_{-1,-1}  +\beta_t d^2_{-1,-1}   \right]
    \\
    \mathcal{M}_{-0,-+}^{h} \ & = \ 0
\end{align}
    \item $(-0,--)$
\begin{align}
    \mathcal{M}_{-0,--}^{\gamma} \ & = \ - \frac{2}{3} \ \left[8 \sqrt{2} G_F m_t m_W s_w^2   \beta_W  \right] d_{-1,0}^1 
       \\
    \mathcal{M}_{-0,--}^{Z} \ & = \ - \frac{2}{3} \ \left[2 \sqrt{2} G_F m_t m_W  \left(\frac{3}{2} - 4s_w^2 \right)   \frac{s}{s - M_Z^2}  \beta_W   \right] d_{-1,0}^1 
    \\
  \mathcal{M}_{-0,--}^{b} \ & = \   \frac{1}{ \sqrt{2}} G_F m_t m_W   \frac{s}{t}   \left[ \left(  \beta_t \beta_W + \beta_t - \beta_W (1 - \beta_W) \right) d_{-1,0}^1  + \frac{2 }{\sqrt{3}} \beta_t  d_{-1,0}^2  \right]   \\
    \mathcal{M}_{-0,--}^{h} \ & = \ 0
\end{align}
\end{itemize}

The corresponding amplitude for $h_{W^+} = 0$, $h_{W^-} = +$ can be found be CP transforming the above expressions.

\begin{itemize}
    \item $(+0,++)$
\begin{align}
    \mathcal{M}_{+0,++}^{\gamma} \ & = \  \frac{2}{3} \left( 8 \sqrt{2} G_F m_t m_W s_w^2 \beta_W   \right) d_{1,0}^1 
       \\
    \mathcal{M}_{+0,++}^{Z} \ & = \  \frac{2}{3} \ \left[2 \sqrt{2} G_F m_t m_W  \left(\frac{3}{2} - 4s_w^2\right) \frac{s}{s - M_Z^2}   \beta_W  \right] d_{1,0}^1  
    \\
 \mathcal{M}_{+0,++}^{b} \ & = \   \frac{1}{ \sqrt{2}} G_F m_t m_W   \frac{s}{t}  \left[\left( \beta_t \beta_W - \beta_t + \beta_W (1 + \beta_W)\right) d^1_{1,0} - \frac{2 }{\sqrt{3}} \beta_t d^2_{1,0}  \right]  \\
    \mathcal{M}_{+0,++}^{h} \ & = \ 0
\end{align}

    \item $(+0,+-)$
\begin{align}
    \mathcal{M}_{+0,+-}^{\gamma} \ & = \  -  \frac{2}{3} \  \left[8 G_Fs_w^2m_W s^{1/2} \beta_W  \right]  d_{1,1}^1  
       \\
    \mathcal{M}_{+0,+-}^{Z} \ & = \ - \frac{2}{3} \ \left[ 2 G_Fm_W \left(\frac{3}{2} \left(1 - \beta_t \right) - 4s_w^2 \right) \frac{s^{3/2}}{s - M_Z^2}  \beta_W \right]  d_{1,1}^1  
    \\
  \mathcal{M}_{+0,+-}^{b} \ & = \ -  \frac{1}{2} G_F m_W \frac{s^{3/2}}{t} (1 -  \beta_t)  \left[ \left( \beta_W(1 + \beta_W) - \beta_t \beta_W \right) d_{1,1}^1  - \beta_t d^2_{1,1} \right] \\
    \mathcal{M}_{+0,+-}^{h} \ & = \ 0
\end{align}
    \item $(+0,-+)$
\begin{align}
    \mathcal{M}_{+0,-+}^{\gamma} \ & = \   \frac{2}{3} \  \left[8 G_Fs_w^2m_W s^{1/2} \beta_W \right]  d_{1,-1}^1   
       \\
    \mathcal{M}_{+0,-+}^{Z} \ & = \  \frac{2}{3} \ \left[ 2 G_Fm_W \left(\frac{3}{2} \left(1 + \beta_t\right) - 4s_w^2 \right) \frac{s^{3/2}}{s - M_Z^2} \beta_W \right]  d_{1,-1}^1  
    \\
    \mathcal{M}_{+0,-+}^{b} \ & = \  \frac{1}{2}  G_F m_W  \frac{s^{3/2}}{t} (1+ \beta_t) \left[ \left( \beta_W(1 + \beta_W) + \beta_t \beta_W \right) d_{1,-1}^1  - \beta_t d^2_{1,-1} \right] \\
    \mathcal{M}_{+0,-+}^{h} \ & = \ 0
\end{align}
    \item $(+0,--)$
\begin{align}
    \mathcal{M}_{+0,--}^{\gamma} \ & = \ - \frac{2}{3} \ [8\sqrt{2} G_F m_t m_W s_w^2  \beta_W   ]  d_{1,0}^1  
       \\
    \mathcal{M}_{+0,--}^{Z} \ & = \  - \frac{2}{3} \ \left[ 2 \sqrt{2} G_F m_t m_W \frac{s}{s - M_Z^2}   \left(\frac{3}{2} - 4s_w^2\right) \beta_W  \right]  d_{1,0}^1  
    \\
    \mathcal{M}_{+0,--}^{b} \ & = \ - \frac{1}{ \sqrt{2}} G_F m_t m_W   \frac{s}{t}   \left[\left( - \beta_t \beta_W + \beta_t + \beta_W (1 + \beta_W)\right) d^1_{1,0} - \frac{2 }{\sqrt{3}} \beta_t d^2_{1,0}  \right]
    \\
    \mathcal{M}_{+0,--}^{h} \ & = \ 0
\end{align}
\end{itemize}
Amplitude for $h_{W^+} = 0$, $h_{W^-} = -$ are related by CP transformation.

\begin{itemize}
    \item $(++,++)$
\begin{align}
    \mathcal{M}_{++,++}^{\gamma} \ & = \   - \frac{2}{3} \ \left[ 8 \sqrt{2} G_F m_t m_W^2 s_w^2 s^{-1/2} \beta_W   \right]  d_{0,0}^1  
       \\
    \mathcal{M}_{++,++}^{Z} \ & = \ -  \frac{2}{3} \ \left[ 2 \sqrt{2} G_F m_t m_W^2 \left(\frac{3}{2} - 4s_w^2 \right)   \frac{s^{1/2}}{s - M_Z^2}    \beta_W \right]  d_{0,0}^1  
    \\
    \mathcal{M}_{++,++}^{b} \ & = \     2 \sqrt{2} G_F m_t m_W^2   \frac{s^{1/2}}{t}  \left[ \frac{1}{6} \left(\beta_t - 3 \beta_W \right) + \frac{d_{0,0}^1  }{2} (\beta_t - \beta_W) + \frac{d_{0,0}^2  }{3} \beta_t\right]
    \\
    \mathcal{M}_{++,++}^{h} \ & = \ 2  G_F \beta_t y_t v m_W^2 \frac{\sqrt{s}}{s - m_H^2}
\end{align}

    \item $(++,+-)$
\begin{align}
    \mathcal{M}_{++,+-}^{\gamma} \ & = \   \frac{2}{3} \ \left[ 8 G_F  m_W^2 s_w^2 \beta_W    \right]  d_{0,1}^1  
       \\
    \mathcal{M}_{++,+-}^{Z} \ & = \   \frac{2}{3} \ \left[ 2 G_F m_W^2 \left(\frac{3}{2} \left(1 - \beta_t \right) - 4s_w^2 \right)  \frac{s}{s - M_Z^2}  \beta_W \right]  d_{0,1}^1  
    \\
    \mathcal{M}_{++,+-}^{b} \ & = \    G_F m_W^2   \frac{s}{t} (1 - \beta_t)  \left[ \beta_W d_{0,1}^1  - \sqrt{\frac{1}{3} } \beta_t d_{0,1}^2   \right]
    \\
    \mathcal{M}_{++,+-}^{h} \ & = \ 0
\end{align}
    \item $(++,-+)$
\begin{align}
    \mathcal{M}_{++,-+}^{\gamma} \ & = \  - \frac{2}{3} \ \left[8 G_F  m_W^2 s_w^2 \beta_W   \right ]  d_{0,-1}^1  
       \\
    \mathcal{M}_{++,-+}^{Z} \ & = \ -  \frac{2}{3} \ \left[ 2 G_F m_W^2 \left(\frac{3}{2} \left(1 + \beta_t \right) - 4s_w^2 \right)  \frac{s}{s - M_Z^2}   \beta_W \right] d_{0,-1}^1  
    \\
    \mathcal{M}_{++,-+}^{b} \ & = \ -  G_F m_W^2  \frac{s}{t} \left(1 + \beta_t \right) \left[ \beta_W d_{0,1}^1  - \sqrt{\frac{1}{3} } \beta_t d_{0,1}^2   \right]
    \\
    \mathcal{M}_{++,-+}^{h} \ & = \ 0
\end{align}
    \item $(++,--)$
\begin{align}
    \mathcal{M}_{++,--}^{\gamma} \ & = \   \frac{2}{3} \ \left[8 \sqrt{2} G_F m_t  m_W^2 s_w^2  s^{-1/2} \beta_W \right]   d_{0,0}^1  
       \\
    \mathcal{M}_{++,--}^{Z} \ & = \    \frac{2}{3} \ \left[ 2 \sqrt{2} G_F m_t m_W^2  \left(\frac{3}{2} - 4s_w^2 \right)  \frac{s^{1/2}}{s - M_Z^2}  
 \beta_W  \right] d_{0,0}^1  
    \\
    \mathcal{M}_{++,--}^{b} \ & = \ - 2 \sqrt{2} G_F m_t m_W^2   \frac{s^{1/2}}{t}  \left[ \frac{1}{6} \left(\beta_t + 3 \beta_W \right) - \frac{d_{0,0}^1  }{2} (\beta_t + \beta_W) + \frac{d_{0,0}^2  }{3} \beta_t\right]
    \\
    \mathcal{M}_{++,--}^{h} \ & = \ -  {2} G_F \beta_t y_t v m_W^2 \frac{\sqrt{s}}{s - m_H^2}
\end{align}
\end{itemize}
Amplitude for $h_{W^+} = -$, $h_{W^-} = -$ are related by CP transformation.

\begin{itemize}
    \item $(+-,++)$
\begin{align}
    \mathcal{M}_{+-,++}^{\gamma} \ & = \   0
       \\
    \mathcal{M}_{+-,++}^{Z} \ & = \ 0
    \\
    \mathcal{M}_{+-,++}^{b} \ & = \    \frac{4}{\sqrt{3}} G_F m_t m_W^2    \frac{s^{1/2}}{t} \beta_t  d_{2,0}^2  
    \\
    \mathcal{M}_{+-,++}^{h} \ & = \ 0
\end{align}

    \item $(+-,+-)$
\begin{align}
    \mathcal{M}_{+-,+-}^{\gamma} \ & = \ 0
       \\
    \mathcal{M}_{+-,+-}^{Z} \ & = \   0
    \\
    \mathcal{M}_{+-,+-}^{b} \ & = \ -  \sqrt{2} G_F m_W^2   \frac{s}{t}    \beta_t (1- \beta_t)  d_{2,1}^2
    \\
    \mathcal{M}_{+-,+-}^{h} \ & = \ 0
\end{align}
    \item $(+-,-+)$
\begin{align}
    \mathcal{M}_{+-,-+}^{\gamma} \ & = \  0
       \\
    \mathcal{M}_{+-,-+}^{Z} \ & = \  0
    \\
    \mathcal{M}_{+-,-+}^{b} \ & = \ \sqrt{2} G_F m_W^2   \frac{s}{t}   \beta_t (1+ \beta_t)  d_{2,-1}^2
    \\
    \mathcal{M}_{+-,-+}^{h} \ & = \ 0
\end{align}
    \item $(+-,--)$
\begin{align}
    \mathcal{M}_{+-,--}^{\gamma} \ & = \ 0
       \\
    \mathcal{M}_{+-,--}^{Z} \ & = \   0
    \\
    \mathcal{M}_{+-,--}^{b} \ & = \  - \frac{4}{\sqrt{3}} G_F m_t m_W^2     \frac{s^{1/2}}{t} \beta_t d_{2,0}^2  
    \\
    \mathcal{M}_{+-,--}^{h} \ & = \ 0
\end{align}
\end{itemize}
Amplitude for $h_{W^+} = -$, $h_{W^-} = +$ are related by CP transformation.

\section{Signal vs Background}

In this section, we show the dependence of the significance on the angular variable $\cos\theta$ and the top quark invariant mass $m_{t \bar{t} }$ in the center-of-mass frame for both the 3 TeV and 10 TeV colliders respectively. Polar angle $\theta$ is defined in the partonic frame and is the angle between $t$-quark and $W^+$ boson.

\begin{figure}[!ht] 
  \centering
        \includegraphics[width=0.48\textwidth]{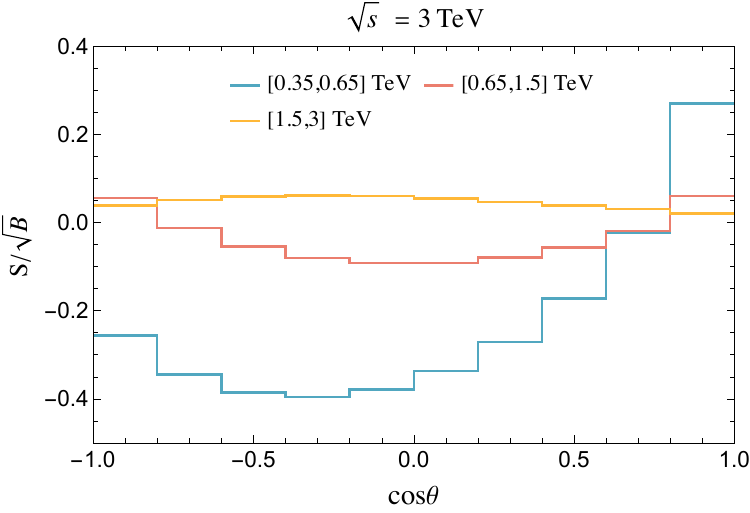}
         \includegraphics[width=0.48\textwidth]{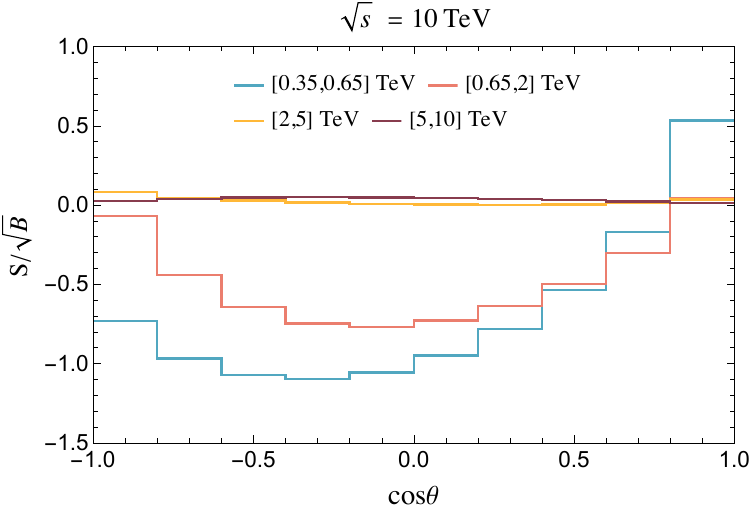}
    \caption{ The significance for 3 TeV (left) and 10 TeV (right) MuC at various ranges of $m_{t\bar t}$ as a function of polar angle. We chose $\delta_{yt} = 0.1$ for 3 TeV collider and $\delta_{yt} = 0.015$ for 10 TeV collider which is near the one sigma C.L. limit for each collider. }
   \label{fig:ang-overlay-10TeV}
\end{figure}

We show angular dependence of signal significance for a $\sqrt{s}$ = 10 TeV and $\mathcal{L}_{int}$ = 10 $ab^{-1}$ collider and for a $\sqrt{s}$ = 3 TeV and $\mathcal{L}_{int}$ = 1 $ab^{-1}$ collider in \cref{fig:ang-overlay-10TeV} for various $m_{t \bar{t} }$ ranges using 50 GeV bins~\footnote{As different lines correspond to different sizes of the $m_{t\bar t}$ energy, we choose to represent the result in such a normalization.}. In this figure, background $B$ refers to the prediction from SM, and signal $S$ refers to the deviation from SM value due to $\delta_{yt}$. The fact that the significance primarily comes from the interference term is explicit here by noticing higher significance near the threshold. In both figures, one can notice that the sign of significance near the forward region is opposite to the rest for lower energy ranges. This pattern can also be observed in \cref{fig:mtt-overlay-3TeV}. The origin of this change of sign in the angular distribution can be attributed to the change of sign for the helicity amplitude $\mathcal{M}_{00,++}$ and $\mathcal{M}_{00,--}$ at $\cos\theta \approx 0.7$ and $\sqrt{\hat{s}} = 400$ GeV. Since the signal is mostly coming from the interference between the above two channels with the contribution from the Higgs exchange (which has a flat angular dependence), the signal follows the shape of $\mathcal{M}_{00,++}$, $\mathcal{M}_{00,--}$ and this can explain why we observe the pattern in \cref{fig:ang-overlay-10TeV}. At higher energies, we will have the contribution from quadratic $\delta_{yt}^2$ term, which is always positive, and so the pattern is not explicit in this region. Another point to note here is that the significance mainly comes from the central bins in angular distribution, so angle cuts in the forward region do not significantly affect our $\Delta \chi^2$ analysis.

\section{Parameter Constraints on VLQ}
The VLQ model modifies the couplings between the top quark and the gauge bosons. The modified Higgs-top coupling would make the precision Higgs observable deviate from the SM predictions, such as the Higgs decay modes to diphoton and di-gluon. In SM, the Higgs boson decay into diphoton occurs via the $W$ boson loop and charged fermion loop \cite{Ellis:1975ap, Shifman:1979eb, Marciano:2011gm}, and Higgs decay to di-gluon receives contributions from quark loops \cite{Ellis:1975ap, Georgi:1977gs}. Here we  focus on the $h\rightarrow gg$ channel since it is more sensitive to the shift $s_L^2$ and only considers the $t$ and $T$ loop, since the heavy quark dominates this loop-induced coupling. 
In the VLQ model, we need to consider the $T$-loop contribution and the shift in Yukawa couplings for the quark loops. Hence in total, the amplitude of Higgs to di-gluon mode before QCD correction is
\begin{equation}
    \mathcal{M}_q^{\lambda \lambda' a b} = \srr{  (1-s_L^2) F_f(\beta_t) + s_L^2 F_f(\beta_T) } \dfrac{\alpha_s g  \brr{k_1 \cdot k_2 g^{\mu\nu} - k_1^\mu k_2^\nu} \frac{1}{2} \delta^{ab}\epsilon_\mu(k_1)^{\lambda} \epsilon_\nu(k_2)^{\lambda'} }{2\pi m_W}
\end{equation}
where,
\begin{equation}
    \beta_t = \dfrac{4 m_t^2}{m_h^2}, \quad
    \beta_T = \dfrac{4 m_T^2}{m_h^2},
\end{equation}

\begin{equation}
    \begin{split}
        F_f(\beta) &= -2 \beta \srr{ 1+(1-\beta)f(\beta) },
    \end{split}
\end{equation}
and 
\begin{equation}
    f(\beta) = \arcsin^2 \brr{\beta^{-\frac{1}{2}}}  \quad \textrm{for} \quad  \beta \geq 1 .
\end{equation}
Since the new physics does not change the kinematics, we can compute the excess squared amplitude and constrain the new physics parameters. Using $\delta \Gamma/\Gamma = \delta |\mathcal{M}|^2 / |\mathcal{M}_{\rm SM}|^2$ and inserting the projected precision for the Higgs decay width to di-gluon at future colliders, we can derive the corresponding allowed region in the parameter plane of the VLQ model under consideration.

\end{appendix}

\bibliographystyle{utphys}
\bibliography{references}

\end{document}